\documentclass{WileyMSP-template}
\usepackage{amsmath}
\usepackage{stackengine}

\usepackage{graphicx}
\usepackage{subfig}
\usepackage{nicefrac}
\usepackage[LGRgreek]{mathastext}
\usepackage{textcomp}

\begin{document}

\pagestyle{fancy}
\rhead{\includegraphics[width=2.5cm]{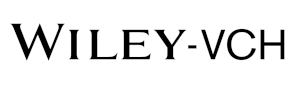}}

\title{A novel receiver design for energy packet-based dispatching}

\maketitle


\author{Friedrich Wiegel*,}
\author{Edoardo De Din,}
\author{Antonello Monti,}
\author{Klaus Wehrle,}
\author{Marc Hiller,}
\author{Martina Zitterbart,}
\author{Veit Hagenmeyer}


\dedication{}

\begin{affiliations}
F. Wiegel, Prof. M. Hiller, Prof. M. Zitterbart, Prof. V. Hagenmeyer\\
Karlsruhe Institute of Technology\\
76131 Karlsruhe, Germany\\
Email Address: friedrich.wiegel@kit.edu
\\[1em]
E. De Din, Prof. A. Monti, Prof. K. Wehrle\\
RWTH Aachen University\\
52062 Aachen, Germany\\






\end{affiliations}



\justifying

\begin{abstract}

A steadily growing share of renewable energies with fluctuating and decentralized generation as well as rising peak loads require novel solutions to ensure the reliability of electricity supply. More specifically, grid stability is endangered by equally relevant line constraints and battery capacity limits. In this light, energy packet-based dispatching with power signal dual modulation has recently been introduced as an innovative solution. However, this approach assumes a central synchronicity provision unit for energy packet dispatching. In order to overcome this assumption, the present paper's main contribution is a design of an energy packet receiver which recovers the required synchronicity information directly from the received signal itself. Key implementation aspects are discussed in detail. By means of a DC grid example, simulation results show the performance and applicability of the proposed novel receiver for packet-based energy dispatching. 
\end{abstract}



\section{Introduction}
\label{sec:1}

Today's electricity grid is based on the equilibrium principle, i.e. the amount of power generated corresponds to the power consumed at every instant of time. Due to the steadily growing number of renewable energies with fluctuating and decentralized generation as well as increasing peak loads due to the rising use of electric vehicles and heat pumps in the context of the energy transition, it is increasingly challenging to guarantee this instantaneous balance while considering local constraints. Demand Response (DR), Demand Side Management (DSM) and Energy Management Systems (EMS) can be considered as potential solutions to this challenge, at least in the short term \cite{STRBAC20084419} \cite{4523466} \cite{ZHOU201630}. Unfortunately, the possibilities for load shifting in today's power grids are limited due to equally relevant line constraints and battery capacity limits. For a long-term solution, it is therefore necessary to consider completely new concepts, which operate without overall synchronicity between generation and consumption.


A packet-based energy distribution concept, as presented e.g. in \cite{8245741} and \cite{Din.2018}, can be one of such long-term solutions. According to the number of recent publications e.g. \cite{Chen.2020}, \cite{9169983} and \cite{Zhang.2020}, such a packet-based energy distribution approach is highly topical and widely accepted. Inspired by the Internet protocol (IP) known from modern communication technologies, an asynchronous energy distribution is a promising option to solve the problem described above and to make the entire power distribution network future-proof. Equivalent to the data packet from the Internet protocol, an energy packet (EP) is defined as the key element in this concept. An energy packet itself is a defined amount of energy that is exchanged within the distribution network between source and sink asynchronously according to demand or availability. In combination with energy storage elements, this approach enables energy transmission based on the store-and-forward principle known from the IP protocol. The two characteristics of this concept, namely the energy packet definition and the store-and-forward principle, result in two major advantages which contribute significantly to the improvement of the energy supply infrastructure: (i) creating a clear decoupling between demand and generation; (ii) always allowing the best use of the infrastructure capacity, i.e. line limits and storage capacity.

While in \cite{8245741} the vision and the underlying theoretical considerations are presented in detail, in \cite{Din.2018} a possible solution for the realization of energy packets and their transmission via a direct current (DC) grid is presented. Thereby, for successful transmission of energy packets, synchronicity on the data level is assumed.

The present contribution significantly extends \cite{Din.2018} by designing the solution of a practicable receiver for energy packets. The synchronization necessary for the detection of the information part of the energy packet is derived from the received packet itself. This means that the previous central synchronization prerequisite is no longer necessary and the energy packets can therefore be exchanged asynchronously over the distribution network, as demonstrated by the simulation results in section~\ref{sec:5}.

This paper is organized as follows: Section~\ref{sec:2} introduces the concept of packet-based energy distribution and outlines the basic idea of power signal dual modulation (PSDM) for packet-based energy distribution. Furthermore, the advantages and disadvantages of related solutions are discussed. In section~\ref{sec:3}, the basic idea of a PSDM transmitter is outlined. Section~\ref{sec:4} describes in detail which steps have to be taken in the communication part of the PSDM receiver in order to transmit energy packets without global synchronicity and how the required point-to-point synchronization between the participants can be achieved from the received signal. In section~\ref{sec:5}, the simulation results are summarized. Finally, the main conclusions and an outlook are provided in section~\ref{sec:6}.
\section{Packet-based energy distribution in DC networks}
\label{sec:2}

\subsection {Concept}

\begin{figure}[tbp!]
  \begin{center}
  \includegraphics[width=4in]{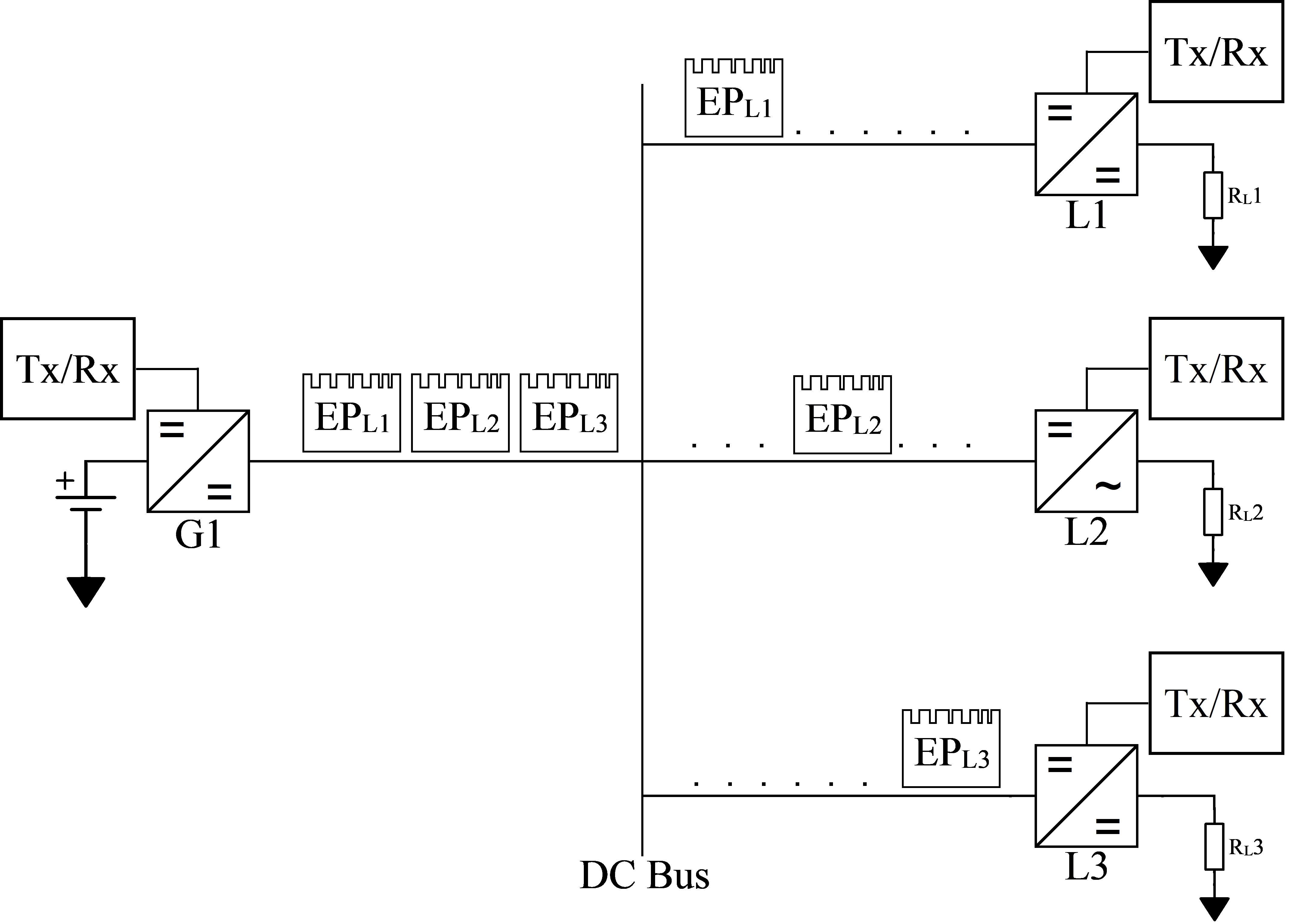}\\
  \caption{Concept of packet-based energy distribution.}
  \label{concept_pbed}
  \end{center}
\end{figure}

\textbf{Figure~\ref{concept_pbed}} visualizes the concept of packet-based energy distribution in a simplified form. Each participant of this simple energy distribution network is connected to a DC bus either via DC/DC (see G1, L1 and L3) or DC/AC converters (see L2). The generator G1 supplies the DC bus with a constant DC voltage $V_{G1}$. Communication paths are represented by transceiver blocks Tx/Rx. All loads are in standby at initial time $t_0$, i.e. the respective input impedance of the converters is very high and no current flows. At time $t_1$, an energy packet is transmitted to the load $L_3$. For this purpose, the load $L_3$ is informed via a communication path that it should switch to the current consumption mode for the duration $T_{L3}$. For the duration $T_{L3}$ a current $I_{L3}$ flows. This process could be interpreted as an energy packet transmission procedure:

\begin{equation}\label{eq:1}
EP_{L3} = v(t)_{G1} \cdot i(t)_{L3} \cdot t_{L3}
\end{equation}

Obviously, the same principles apply to the transmission of $EP_{L2}$ at $t_2$ and $EP_{L1}$ at $t_3$. Taking the concept shown in Figure~\ref{concept_pbed} as a basis, it becomes clear that an energy packet is a combination of the energy flux and the data necessary to transmit an energy packet to the destination load.\footnote{Note that the power part of the energy packet transceiver does not necessarily have to be implemented as a DC system. AC systems with fixed or variable frequency are also conceivable. The only requirement is an inverter suitable for the application needs. The exemplary specification on a DC bus presented in the following is only due to the coherence to the previous publications.} The minimum data are the address of the destination and the size and duration of the energy packet. In a more advanced network, where several intermediate transmissions between the source and the destination may be necessary, additional information such as the address of the intermediate destination, number of hops, etc. may also be required.

\subsection {Energy packets metadata providing strategies}

The information path for the control of the  energy packet transmission mentioned above can be realized in various ways. The use of industrial or conventional wired communication technologies such as EtherCAT or IEEE 802.3(xx) is conceivable. However, this approach results in significant additional installation and cost effort and is therefore not practical in every section of an energy distribution network.

Wireless low-rate communication technologies such as IEEE 802.15.4, Z-Wave or EnOcean for low-rate wireless personal area networks (LR-WPAN) could be a possible solution for dedicated power distribution network subsections such as a house, an industrial building or an industrial plant \cite{Sato.2015} \cite{Gungor.2010} \cite{Tuna.2013}. These are known among others from several applications in the field of Smart Homes and Home Automation and are becoming more and more popular. These communication technologies provide sufficient high bandwidths for the applications mentioned, have low production costs and enable the plug-and-play expansion of the communication network without installing additional cables.

As described in \cite{Wiegel.81220188152018}, all these low-cost and low-power radio communication standards unfortunately also have some critical disadvantages that make the use of these techniques problematic in the context of energy packet transmission. Due to the very low transmission power defined by the respective standards and the cost-effective design of the transceivers, the established network is very sensitive to any kind of interference and multipath influences, which are always present in buildings and industrial plants \cite{.2002}. Therefore, a network created by these communication techniques is called a Lossy Network, i.e. a high number of lost information packets is expected by definition.  This is also reflected in the name of the Routing Protocol for low power and lossy Networks (RPL) \cite{RPL}. Due to the desired high degree of reliability of the power distribution network, lossy networks are not appropriate in this case. In addition, the coverage of such networks is relatively small \cite{.2002} \cite{Wehrle.2010}. Moreover, building topologies such as steel concrete ceilings or electromagnetic shielding are challenging since they can lead to non-accessibility of net-elements by radio. Especially many large loads or generators are installed in rooms with electromagnetically shielding walls. Even in industrial plants, large loads or generators are often not easily accessible by radio.

However, in most cases a power outlet is within reach. Since all participants of the power distribution network have to be connected to the powerline, powerline communication (PLC) is a natural choice for the transmission of the control data at least within certain subsections with a limited extension and a relatively manageable number of participants.

The Powerline communication techniques themselves can be divided in two groups: Broad Band - Powerline Communication (BB-PLC) and Narrow Band - Powerline Communication (NB-PLC). In Europe, BB-PLC communications technology operates in a frequency band between 1.6 MHz and 30 MHz and can provide data rates of several gigabit per second (Gbps) \cite{Cano.2016}. However, since these data rates can only be achieved under optimal conditions and are not necessary for the transmission of energy packets, and the BB-PLC transceivers are significantly more expensive in production than NB-PLC transceivers, the use of this technology is not considered. In contrast to BB-PLC, NB-PLC technology operates in a frequency band between 3 kHz and 500 kHz and provides data rates of up to one megabit per second, depending on the communication standard and frequency band. The coverage radius is between hundred meters and several kilometers, depending on the standard used \cite{Cano.2016} \cite{Tonello.2016}. Therefore, NB-PLC is definitely a promising technique in the context of grid element control and smart grid applications, as demonstrated in several studies \cite{Lampe.2011, 6094006, 5768099}. Unfortunately, the European Committee for Electrotechnical Standardization (CENELEC) comparatively limits the permissible frequencies and transmission powers for NB-PLCs in Europe to a very high degree, thus immensely limiting the achievable transmission rates. The European standard (EN) defines the in-band and out-of-band emission limits and states that the CENELEC A band (3-95 kHz) is reserved for energy suppliers and that only the CENELEC B-D bands (95-148.5 kHz) may be used by consumer installations \cite{EN50065}. Due to these restrictions, conventional NB-PLC technology operates in Europe exactly within a frequency band where the power electronics required for generating the energy packets have the maximum disturbance emission, and consequently, communication is vulnerable to the switching frequency noise.
 
\subsection {Power Signal Dual Modulation based Powerline Communication for Energy Packet distribution}

To overcome this problem, powerline communication based on Power-Signal-Dual-Modulation (PSDM) was proposed in the articles \cite{Wu.2015}, \cite{Wang.2017}, \cite{Du.2017} and \cite{Din.2018}.  With PSDM, the information signal is embedded into the power signal by manipulating the pulse width modulation signal of the power converter. Either the phase or the frequency of the PWM signal can be manipulated to inject the information. In this way, both data modulation and power conversion are implemented in a single electronic circuit and no explicit analog front ends or coupling units are required like in conventional PLC transmitters. This simplifies the system structure and minimizes implementation costs.

The paper \cite{Din.2018}, on which the present contribution is based on, describes exactly how energy packets can be generated and transmitted using PSDM technology. However, the paper does not answer how the synchronization required for processing the information part of an energy packet can be achieved. Other publications dealing with PSDM also do not give a satisfactory answer to the  problem of synchronization in the receiver. For example, the theoretical demodulation process of a DBPSK is described in the \cite{Zhu.2018} and it is also pointed out that the symbol clock is crucial for the demodulation of the data, but at the same time it is written: "Still, bit-synchronization is required, which can be easily achieved by program". However, as described in section \ref{sec:4} bit-synchronization or more precisely symbol-clock recovery is not an easy procedure at all, and is essential for processing and interpreting the received information. In \cite{Wang.2017} a PSK/DSSS modulated signal with suppressed carrier is coherently demodulated. The authors noted that accurate information about the frequency and phase of the carrier is necessary to demodulate the data. To obtain this information, they propose to transform the received PSK/DSSS signal simply into the frequency domain without any pre-operation. However, a non-linear operation has to be applied to a suppressed carrier information signal first to obtain a clear spectral line of the carrier, as can be found in \cite{Johnson.2000} or \cite{Anderson.2006}.

Furthermore, another essential component of a receiving unit, the Adaptive Gain Control, has been completely disregarded in all cited articles, whereby in all of them the interpretation of the data depends on a normalized level. 

Hence, the present contribution discusses possible solutions for synchronization and level stabilization in the information path in the context of packet-based energy distribution based on PSDM.


\section{PSDM Transmitter for Energy Packet generation}
\label{sec:3}

\begin{figure}[tbp!]
  \begin{center}
  \includegraphics[width=3.45in]{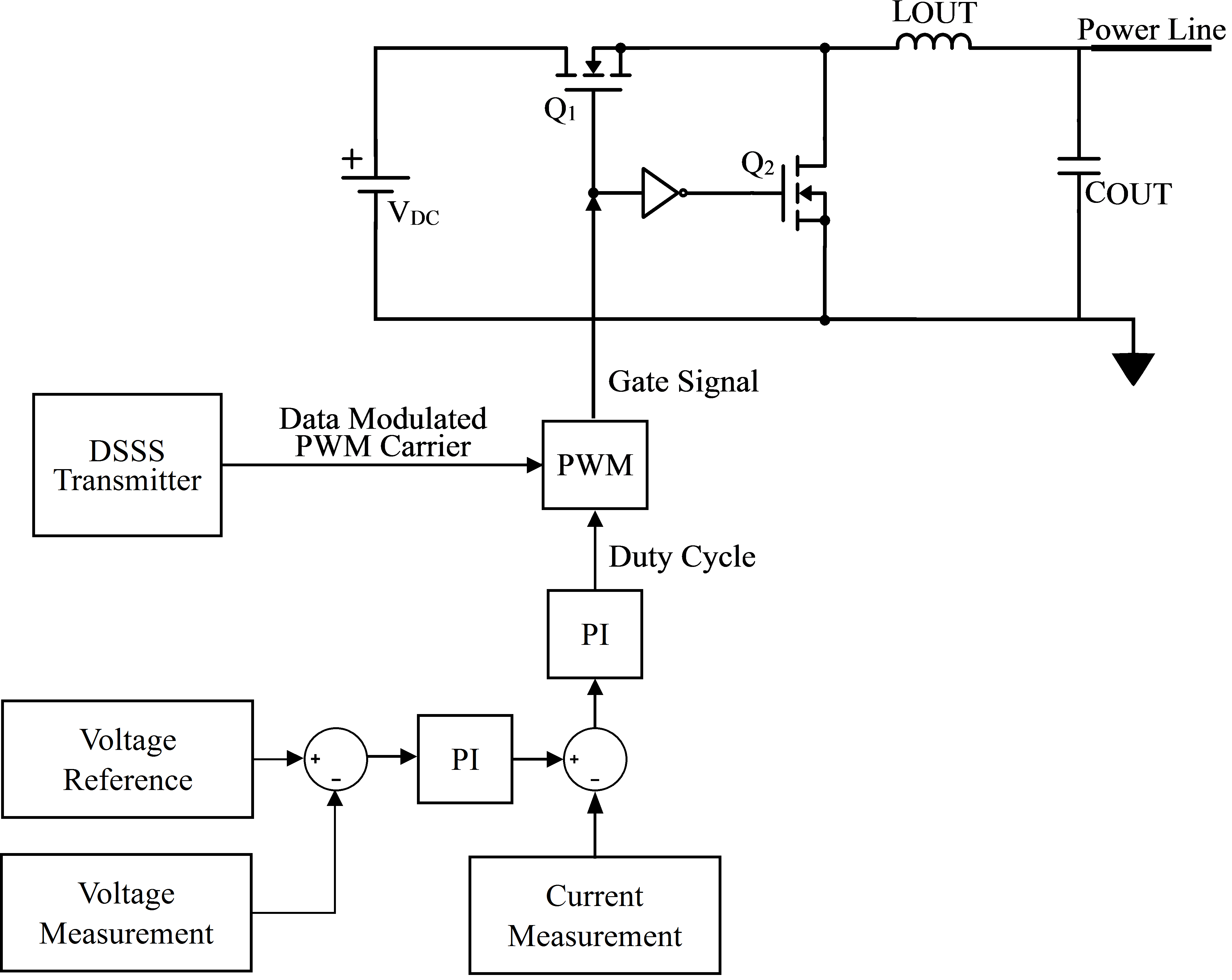}\\
  \caption{Block diagram Energy Packets Source}
  \label{block_diagramm_eps}
  \end{center}
\end{figure}

To provide an input signal for the implemented receiver, a power signal dual modulation based generator for energy packets as proposed in \cite{Din.2018} is realized. \textbf{Figure~\ref{block_diagramm_eps}}  shows the structure of this generator. As the name implies, the generator performs two tasks: First, it acts as a buck converter, which supplies the DC bus with a constant DC voltage and the energy during the energy packet transmission phase. At the same time, it also acts as an analog part of a powerline transmitter by injecting the information signal into the power line without additional circuits such as an analog front end or coupler. In the following simulation, the output signal of this generator serves as input signal for the realized receiver.

As shown in Figure~\ref{block_diagramm_eps}, the information signal is embedded into the power signal by manipulating the duty cycle of the PWM signal used to control the power transistors. The theoretical basics of this procedure are explained in detail in \cite{Din.2018} or \cite{Wang.2017}. 

The transmitter responsible for generating the modulated PWM carrier is based on the Direct Sequence Spread Spectrum (DSSS) technology. For this purpose, the data to be transmitted are first DBPSK-coded, then spreaded with an orthogonal pseudo-random sequence based on the Walsh-Hadamard matrix and finally modulated onto the PWM carrier. The use of DSSS transmission technology makes it possible to secure the data connection against the pulse interference and multi-path propagation strongly represented within the power line and thus to design the control path reliably \cite{Viterbi.2001} \cite{Han2017_1000068355}. It also ensures a certain level of security, since only the load that knows the corresponding despreading sequence can decode the data. 

\section{Receiver for the information part of Energy Packets}
\label{sec:4}

\begin{figure}[ptbh!]
  \begin{center}
  \includegraphics[width=3.45in]{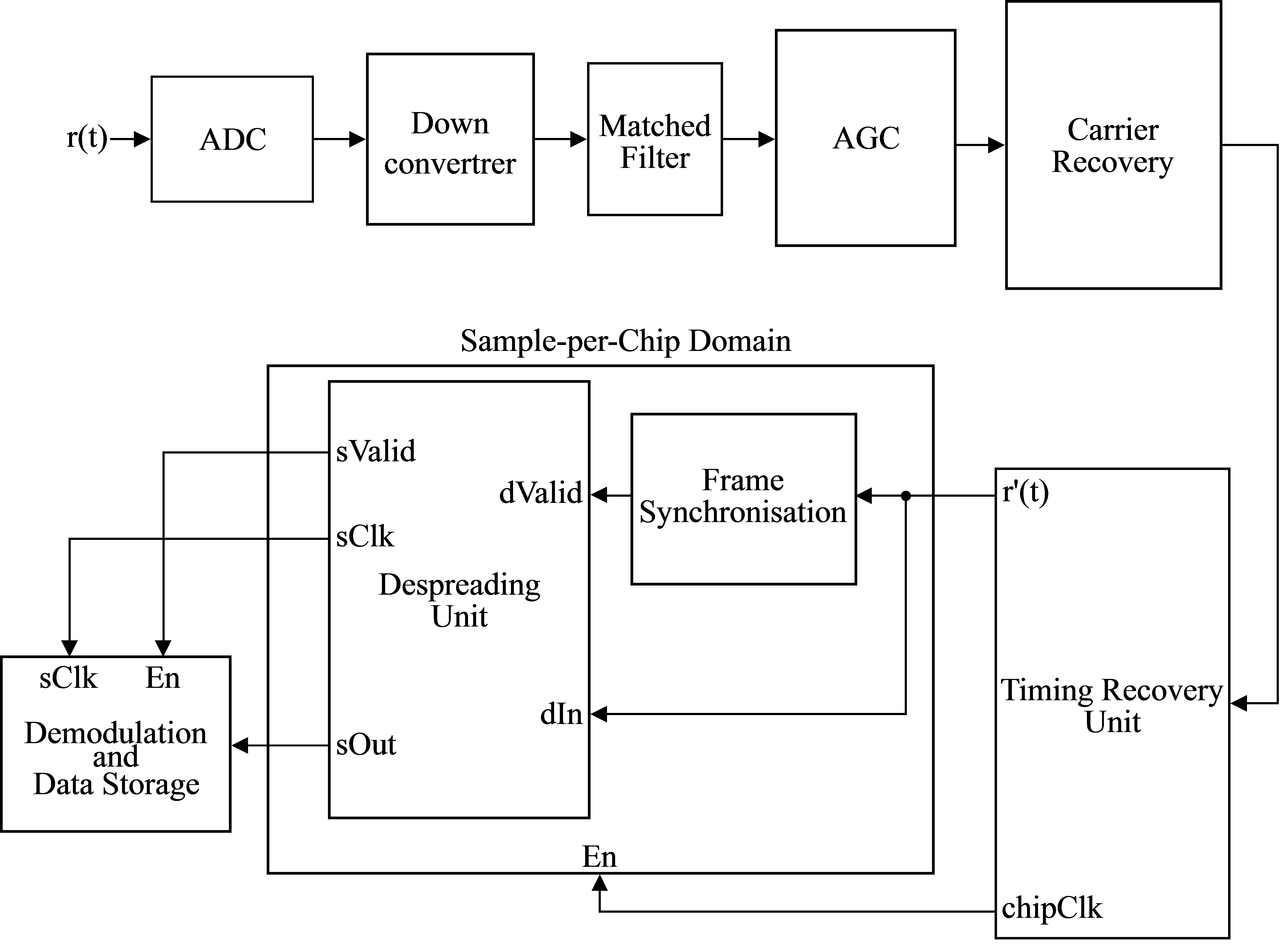}\\
  \caption{Block diagram PLC receiver}
  \label{block_diagramm_plc_rx}
  \end{center}
\end{figure}

\textbf{Figure~\ref{block_diagramm_plc_rx}} shows the implemented powerline communication receiver. In principle, every receiver and also the implemented powerline communication receiver is a transmitter backwards that contains some additional elements, especially several indispensable synchronization units. These are necessary because the receiver generally has no precise information about the symbol, i.e. chip clock, frequency and phase of the carrier and the start point of the data packet. Furthermore, the receiver has to derive this information from the received signal itself. In addition, the clock elements in both the transmitter and receiver have a certain temperature drift, manufacturing tolerances, ageing phenomena, which cause a variable time offset \cite{article.Zhou}.

The resulting time errors have an influence on the signal level within the digital signal processing and can significantly influence the reception quality or even make data transmission impossible \cite{Anderson.2006}. Therefore, the receiver has to perform at least the following synchronization tasks for a successful data transfer  \cite{Anderson.2006} \cite{iq5375.c} \cite{Johnson.2000}:
\begin{itemize}
    \item Carrier synchronization
    \item Timing recovery
    \item Frame synchronization
\end{itemize}

In general, two additional synchronization stages are required for direct spread spectrum systems:

\begin{itemize}
    \item Code acquisition
    \item Code tracking
\end{itemize}

However, these are only indispensable if one has to deal with strong frequency-selective interference or if one wants to implement a system with code multiplexing media access. However, since we primarily use spread spectrum technique to assign the messages to the respective load and our intention is to show that the synchronization approaches presented afterwards can also be used for power/signal dual modulation, we can skip with these two synchronization stages in the first step. This is possible because a DBPSK/DSSS signal can be treated for synchronization purposes as a pure DBPSK signal with a data rate increased by the spreading factor. Furthermore, the synchronization approaches described in the following are so generically implemented that they can be modified to a strict DSSS solution with a manageable effort. Among the required adjustments are the replacement of the Timing-Error Detector (TED) by a code phase error detector, the addition of a code acquisition unit and the calculation of the carrier phase error using the despread symbols and not the baseband signal. Thus, the solutions shown in this paper can be considered as the basis for further synchronization steps.

Next, the realized synchronization levels \emph{4.1. Adaptive Gain Control} - \emph{4.5. Frame Synchronisation} are explained in detail. The corresponding simulation results follow in Section~\ref{sec:5}. The description of the general elements of the receiver, such as matched filter, demodulation unit, etc., which form the counterpart to the transmitter, is omitted, but reference is made to further literature where these elements are described in detail \cite{Johnson.2000} \cite{Anderson.2006} \cite{Rice_2009}.

\subsection{Adaptive Gain Control}
\label{AGC}

Phase Locked Loop (PLL) based synchronization algorithms require a constant average signal energy, since this influences the $K_p$ factor of the Phase Error Detector or Timing Error Detector \cite{Rice_2009}. The level of the threshold value for frame synchronization also depends on the average signal energy of the incoming signal \cite{1094813}. Thus, a constant signal level is a critical parameter for the realization of a digital transmission system. This is ensured by the Adaptive Gain Control (AGC) unit.

As shown in Figure~\ref{block_diagramm_plc_rx}, the signal level correction via AGC takes place in the baseband, i.e. after the down-conversion of the input signal, but before any synchronization. At this point the signal can be described by:

\begin{equation}\label{eq:2}
u(k)= A(k) \cdot e^{j\phi_r(k)}
\end{equation}

where $u(k)$ represents the input signal, $A(k)$ the amplitude of the symbol and $\phi_r(k)$ the phase of received symbols. Due to the variable distance between the transmitter and receiver, the used transmission power as well as different attenuation influences within the channel, the amplitude $A(k)$ of the signal is variable and even time-variant. In order to compensate this variance and to adjust the signal level to the reference level, the AGC unit is used. Following difference  equations describe the implemented Least Mean Squares (LMS) signal level adaptation algorithm:


\begin{align}
\label{eq:3}
    x(k+1) ={}& x(k) \cdot \left(1-\alpha|u(k)| \right)+\alpha R \\
\label{eq:4}
    y(k) ={}& x(k) \cdot u(k)
\end{align}

where

\addstackgap[5pt]{\begin{tabular}{lll}
    $\bullet$ & $u(k)$ & represents the complex input signal\\
    $\bullet$  & $R$ & represents the reference signal level\\
    $\bullet$  & $\alpha$ & represents the step size\\
    $\bullet$  & $x(k)$ & represents the divide by factor\\
    $\bullet$  & $y(k)$ & represents the output signal of the AGC
\end{tabular}}

To ensure a constant signal level, the magnitude of the AGC output signal $y$ is compared with a reference signal level $R$. If the output signal level is too high (low), a negative (positive) signal is fed back, reducing (increasing) the gain. The control parameter $\alpha$ regulates the amplitude of the feedback signal and is used to control the AGC’s time constant.

Corresponding to Eqn. (\ref{eq:3}) the calculation of the absolute value of the complex input signal $|u(k)|$ for determining the amplitude of the input signal is a non-linear process, so the resulting equation is also non-linear. However, based on the assumption that the system is driven by a step $u(k) = c_{k}$ with $c_{k} > 0$, the non-linear equation becomes a linear difference equation:
\begin{equation}\label{eq:5}
x(k+1) = x(k) \cdot (1-\alpha c)+\alpha R
\end{equation}


From this solution follows that in steady state the gain is $x(k) = \frac{R}{c}$ and the system time constant is $\tau = \frac{1}{\alpha c}$.

\subsection{Carrier synchronization}

\begin{figure}[tbph!]
  \begin{center}
  \includegraphics[width=2.75in]{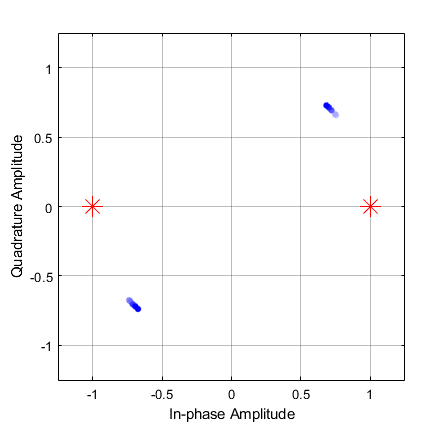}\\
  \caption{Deviation of the symbols from the nominal value when there is a phase offset between the local oscillator and the carrier wave}
  \label{fig:4}
  \end{center}
\end{figure}

A phase deviation of the local oscillator of the down-converter from the carrier wave regardless of whether time-variant or not causes an offset (if time-invariant) or rotation (if time-variant) of the received symbols in the symbol domain \cite{iq5375.c}. \textbf{Figure~\ref{fig:4}} shows an example of such a  time-variant deviation of the received symbols (blue) from the expected symbols (red). At the shown snapshot the phase shift of the symbols is approx. 45\textdegree. If the offset is large enough - in the case of the BPSK more as $\pm 90$\textdegree - a wrong symbol will be detected.

\begin{figure}[tbp!]
  \begin{center}
  \includegraphics[width=3.0in]{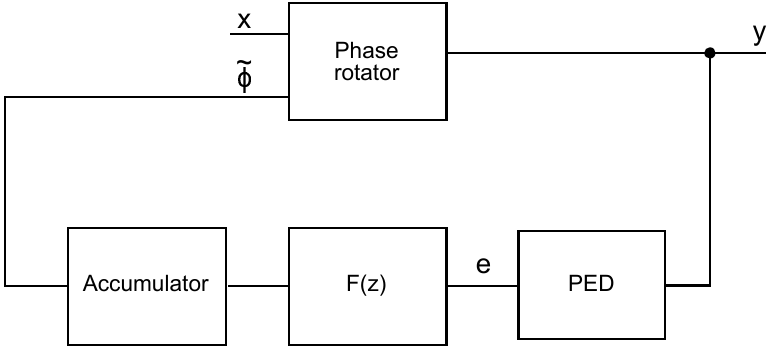}\\
  \caption{Carrier synchronization unit}
  \label{fig:5}
  \end{center}
\end{figure}

The task of carrier synchronization is to keep this phase offset as low as possible. In principle, carrier synchronization is a process of tracking the frequency or phase of the local oscillator to the frequency and phase of the carrier wave. This can be undertaken in different ways \cite{Anderson.2006} \cite{iq5375.c} \cite{Rice_2009}. To avoid the need for additional frequency offset estimation and to keep the complexity of the receiver as low as possible, a PLL-based decision-feedback carrier synchronizer is implemented. This has the advantage that it can correct both phase and frequency offset at least within a limited range. As shown in \textbf{Figure~\ref{fig:5}}, the implemented carrier synchronization unit consists of a carrier phase error detector (PED), a loop filter $F(z)$, a phase error accumulator as well as a phase rotator.

The carrier phase error detector determines the phase error between the carrier wave and the local oscillator. The loop filter removes the unwanted high-frequency signal components and generates the corresponding control signal for the phase error accumulator. The phase error accumulator is responsible for determining an accumulating phase error. The phase rotator eliminates the estimated phase error.

Since the functionality of the used Proportional–Integral loop filter and phase error accumulator is known from discrete PLLs, only the principle of the used carrier synchronization and the functionality of the implemented carrier phase error detector are presented in the following. For the general functionality of the loop filter and phase error accumulator please refer to the further literature \cite{Rice_2009} \cite{Gardner_PLL}.

The received signal can be described by:
\begin{equation}\label{eq:7}
r(t) = G_a\sum_{k} \left\{i(k)p(t-kT_s)\cdot cos(\omega_0t+\varphi_e)\right\}+w(t)
\end{equation}

where the representations are

\addstackgap[5pt]{\begin{tabular}{lll}
    $\bullet$  & $G_a$  & gains and losses of the signal\\
    $\bullet$  & $i(k)$ & k-th chip\\
    $\bullet$  & $p(t)$ & unity-energy pulse shape\\
    $\bullet$  & $\varphi_e$ & unknown combined phase offset\\
    $\bullet$  & $T_s$ & sample period\\
    $\bullet$  & $\omega_0t$ & angular frequency of carrier wave\\
    $\bullet$  & $w(t)$ & addative white Gaussian noise
\end{tabular}}

Without loss of generality and for simplicity, only the part of the overall function that is relevant for carrier synchronization is used in the following:
\begin{equation}\label{eq:8}
r(k) = i(k) \cdot cos(\Omega_0 k+\phi_e)
\end{equation}

The combined phase shift $\phi_e = \Delta \omega t + \Delta \phi$ consists of a possible frequency shift $\Delta \omega$ and a constant phase shift $\Delta \phi$ between the carrier wave and the local oscillator. To eliminate $\phi_e$, an asynchronous shift of the passband signal into the baseband is first performed by means of the down-converter using a complex signal $s(k) = 2 e^{-j\Omega_0k}$, cf. Figure~\ref{block_diagramm_plc_rx}:
\begin{equation}\label{eq:9}
r_{bb}(k) = r(k) \cdot s(k) = i(k) \cdot cos(\Omega_0 k +\phi_r) \cdot 2 e^{-j\Omega_0k}
\end{equation}

The double-frequency components resulting from the multiplication are then eliminated by the matched filter and the following relation is obtained:
\begin{equation}\label{eq:10}
x(k) = i(k) \cdot e^{-j\phi_r} = |i(k)|e^{j\phi_d} \cdot e^{-j\phi_e}
\end{equation}

This means that the symbols $x(k)$ of a signal mixed down asynchronously are shifted in the symbol domain by phase offset $\phi_e$ from the nominal position $\phi_d$ of the symbols as shown in Figure~\ref{fig:4}. The phase error increment that occurs during a single chip interval is calculated in the implemented Phase Error Detector as follows:

Angle of the received symbol:
\begin{equation}\label{eq:11}
\phi_r = tan^{-1} \frac{Im \big( y(k) \big)}{Re \big( y(k) \big)} 
\end{equation}

Angle of the nearest possible symbol :
\begin{equation} \label{eq:12}
\begin{split}
\phi_d & = tan^{-1} \frac{Im \big( y_d(k) \big)}{Re\big( y_d(k) \big)} \\
&= tan^{-1} \frac{0}{Re\big( y_d(k) \big)} \\
&= tan^{-1} \frac{0}{\left | i(k)  \right | \cdot sgn\left[ Re \big( y_r(k) \big) \right]}
\end{split}
\end{equation}

Resulting phase error:
\begin{equation}\label{eq:13}
e(k) = \phi_r - \phi_d = tan^{-1} \frac{Im \big( y(k) \big)}{Re \big( y(k) \big)} - tan^{-1} \frac{0}{\left | i(k)  \right | \cdot sgn\left[ Re \big( y_r(k) \big) \right]}
\end{equation}

\begin{figure}[htbp!]
  \begin{center}
  \includegraphics[width=3.45in]{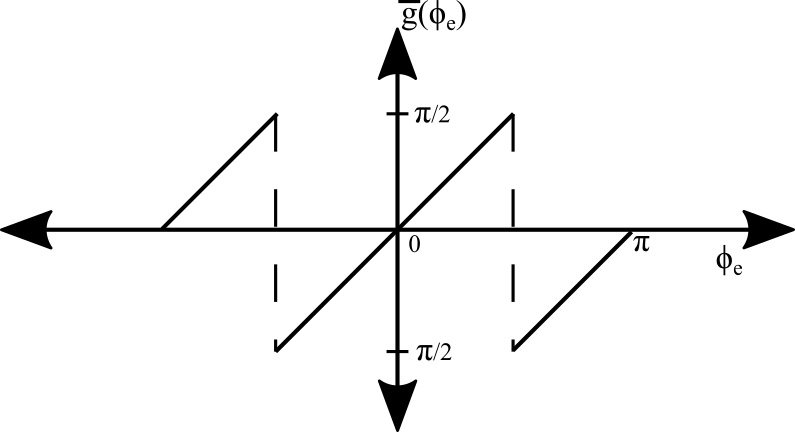}\\
  \caption{S-Curve phase error detector}
  \label{fig:6}
  \end{center}
\end{figure}

The resulting correction phase is then formed using the phase error accumulator by integrating the phase error; subsequently the phase rotator eliminates the offset. \textbf{Figure~\ref{fig:6}} shows the S-Curve of the Phase Error Detector.  Observe that the S-Curve has two stable locking points at $\phi_e = 0$ and $\phi_e = \pm\pi$. Therefore, a phase ambiguity of $\pi$ exists. This disadvantage is eliminated by difference coding in the transmitter. As mentioned in section 3.

\subsection{Timing recovery}

The output of the matched filter must be sampled periodically at the corresponding chip\footnote{A chip is a single elementary modulation state within the Direct Sequence Spread Spectrum. The sequence of different chips is determined by both the spreading code and the transmitted symbols. The chip rate of the spreading code is typically higher than the symbol rate. The ratio between chip rate and symbol rate is called spreading factor SF and is defined as: $SF = \frac{R_c}{R_s} \gg 1$.} rate $f_c$ at time $t_k=kT_c+\tau$, where $T_c$ describes the chip period and $\tau$ the time delay due to the signal propagation between the transmitter and the receiver. Since $\tau$ is generally unknown and $T_c$ in the transmitter and receiver are not completely identical, e.g. due to production tolerances of the clocking elements or thermal variance \cite{article.Zhou}, the optimum sampling point is generally unknown. Not sampling the output of the matched filter at this optimum sampling point $t_k$ and with the locally generated asynchronous sampling frequency $\widetilde{f_c}$ results in periodic degradation of the signal level, leading either to a wrong symbol decision or to temporary full signal loss \cite{Anderson.2006} \cite{iq5375.c}, cf. \textbf{Figure~\ref{fig:s.5_p.7}} in Section \ref{sec:5}.

\begin{figure}[tbp!]
  \begin{center}
  \includegraphics[width=3.45in]{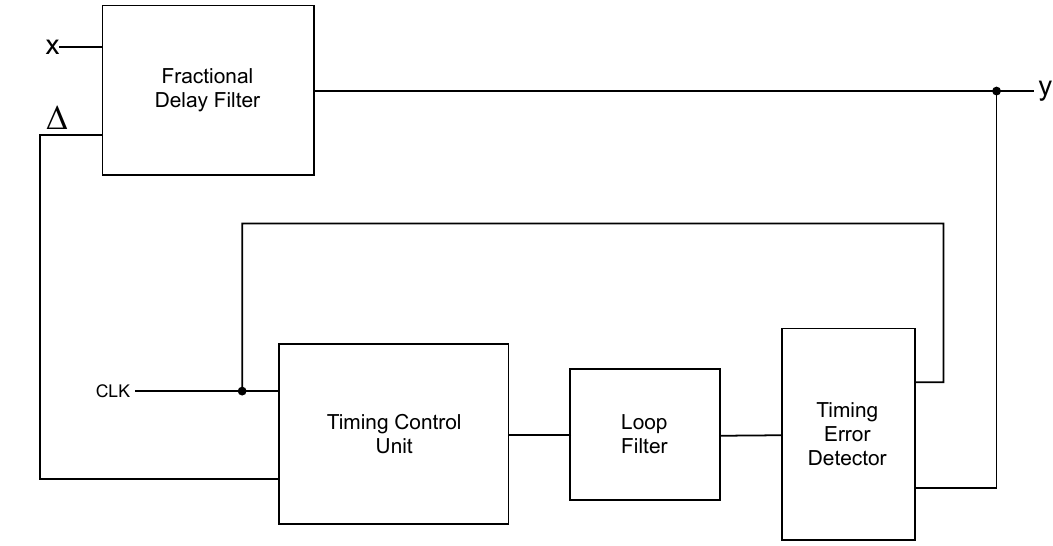}\\
  \caption{Timing Recovery Unit}
  \label{fig:7}
  \end{center}
\end{figure}

The task of the Clock Recovery Unit is to find this optimal sampling point, i.e. to synchronize the receiver with the received signal. Note that since both the installation of a separate clock line between all devices and the reduction of the available signal power (respectively transmission bandwidth) in favor of an explicit clock signal are not viable options, the symbol clock must be derived from the received data. This can be executed both in the time-continuous domain and in the time-discrete domain \cite{Rice_2009}. In the presented article, a time-discrete solution first introduced by Lars Erup and Floyd M. Gardner is implemented \cite{Gardner_I} \cite{Erup.1993}. \textbf{Figure~\ref{fig:7}} shows the block diagram of the implemented clock synchronization unit. This consists of a 3rd order Fractional Delay Filter, a Gardner Timing-Error Detector, a loop filter and a Timing Control Unit.

In the first step, the received continuous signal is oversampled discretized with a fixed asynchronous local frequency $f_s = \nicefrac{1}{T}$. Then, with the aid of the down-converter, it is mixed down into the baseband area and fed to the matched filter. After signal level correction in the AGC unit and carrier synchronization, the signal can be described as follows:

\begin{equation}\label{eq:14}
x(kT) = \sum_{n} i(n) \cdot r_p(kT-nT_c-\tau)
\end{equation}

Thereby $i(n)$ describes the n-th chip, $T$ the sampling period, $T_c$ the chip duration and $r_p(u)$ is the auto-correlation function of the pulse shape and $\tau$ the unknown timing delay. The goal of the Timing Recovery Unit is to eliminate $\tau$, i.e. each resulting sample is aligned with the maximum eye opening. 

This can be achieved by shifting the asynchronous signal by means of fractional delay interpolation, so that after interpolation the data looks as if it had been sampled at the optimal sampling point. To accomplish this, first the timing error $e(n)=f(\tau)$ is determined in the Timing Error Detector. Then $e(n)$ is fed to the loop filter to determine the corresponding phase and frequency error of local chip clock. Subsequently, the Timing Control Unit generates a fractional delay $\Delta$ from the loop filter output signal. This fractional delay $\Delta$ is required to correct the estimated timing offset $\widetilde{\tau}=-\tau$ by sample value correction via interpolation in the interpolator. At the same time, the Timing Control Unit calculates the optimum sampling time in the maximum amplitude of the symbol, i.e. it adjust the local chip clock.

\subsubsection{Interpolation unit}
As mentioned above, the fraction delay filter is used to compute desired samples of $y(nT_{c})$ at the optimum sampling instances from the available sample $x(kT)$. To ensure a linear-phase transmission behaviour of the filter, an odd-order polynomial is chosen. However, due to the fact that interpolation with a 1st degree polynomial leads to large deviations, the next option of a 3rd degree polynomial is used. The resulting algorithm for calculating the desired interpolation value is:

\begin{equation} \label{eq:15}
y(nT_{c}) = x\left ( \left (k+\Delta\right ) T\right ) = \sum_{l=0}^{p} \Delta^{l}\sum_{i=-2}^{1}b_{l}\left ( i \right ) x\left ( \left ( \Delta-i \right ) T \right )
\end{equation}

where $p$ represents the order of the used interpolation polynomial, $\Delta$ the fractional delay of the Timing Control Unit and $b_{l}  \left ( i \right )$ the corresponding filter coefficient, \textbf{Table~\ref{table:1}}.

\begin{table}[]
\begin{center}
\begin{tabular}{r|cccc}
i           & \textbf{$b_3(i)$} & \textbf{$b_2(i)$} & \textbf{$b_1(i)$} & \textbf{$b_0(i)$} \\ \hline
-2          & 1/6              & 0                & -1/6             & 0                \\
-1          & -1/2             & 1/2              & 1                & 0                \\
0           & 1/2              & -1               & -1/2             & 1                \\
1           & -1/6             & 1/2              & -1/3             & 0               
\end{tabular}
\end{center}
\caption{fraction delay filter coefficient}
\label{table:1}
\end{table}

\begin{figure}[htbp!]
  \begin{center}
  \includegraphics[width=3.45in]{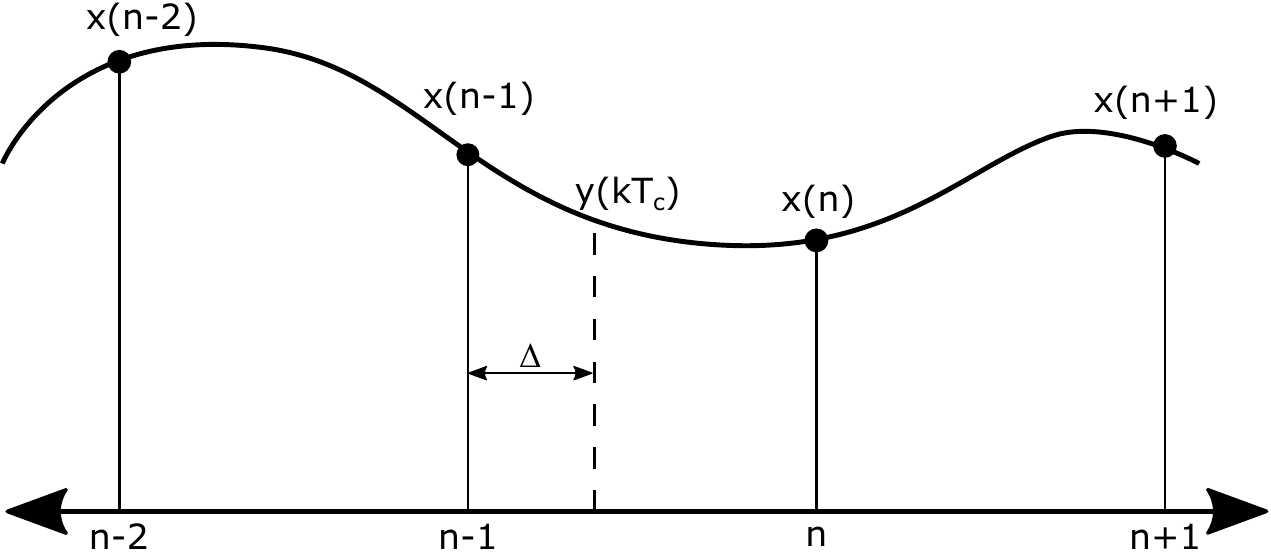}\\
  \caption{Interpolation principle}
  \label{fig:8}
  \end{center}
\end{figure}

These filter coefficients are obtained by determining the polynomial coefficients of the 3rd degree interpolation polynomial: 
\begin{equation} \label{eq:16}
x(k+\Delta) = b_3 \left [ \left (k+\Delta \right)T \right]^3 +  b_2 \left [ \left (k+\Delta \right)T \right]^2 + b_1 \left [ \left (k+\Delta \right)T \right] +  b_0
\end{equation}
as a function of $\Delta$, taking into account the four given samples, which are arranged in pairs to the left and right of the value to be interpolated, cf. \textbf{Figure~\ref{fig:8}}.

A detailed description of the calculation of the filter coefficient as well as a deeper description of the functionality of the Timing Control Unit is omitted and reference is made to the corresponding sources \cite{Rice_2009} \cite{Erup.1993} \cite{Gardner_I}.

\subsubsection{Timing Error Detector}
Implemented TED use the Gardner algorithm to determine one timing error value e(n) per chip. The algorithm is based on a delay difference between the current sample and another sample of the same symbol delayed by half the symbol period and and requires 2 samples per chip. To determine the timing error, 3 consecutive samples are used which are shifted by half a chip to each other. The error signal is finally obtained according to the following equation \cite{Gardner.1986}:

\begin{equation}
\label{eq:17}
e\left ( n \right ) = y  \left ( \left ( n - \nicefrac{1}{2} \right ) T_c \right) \cdot \left [ y \left ( \left ( n - 1 \right ) T_{c} \right ) - y\left ( n T_{c} \right ) \right] 
\end{equation}

Where $y(nT_c)$ are generally complex values. If the sampling is performed at the optimum time point, two values are located in the chip center and one exactly at the transition between two chips, $e(n)=0$. If sampling is premature, then $e(n) < 0$, if sampling is delayed, then $e(n) > 0$.
The advantage of this algorithm is the insensitivity to a carrier frequency offset, thus no previous carrier synchronization is required.

\subsubsection{Timing Control Unit}
The Timing control unit provides the interpolator with the fractional interval $\Delta$ and the TED with the strobe signal CLK to calculate the correct timing errors once per chip.  The interpolation is performed for every sample, and a strobe signal is used to determine if the interpolant is taken as output value. The operation principle of the timing controller is based on the Modulo-1 decrements counter. The counting frequency is determined by the constant $\nicefrac{1}{N}$, where N is the number of samples per chip. The output of the Modulo-1 counter is defined as:

\begin{equation}\label{eq:17}
\Delta(n) = \begin{cases}
N \cdot \delta(k) & \text{CLK = 1} \\
\Delta(n-1) & \text{else}
\end{cases}
\end{equation}

with

\begin{equation}\label{eq:18}
\delta(k) =  \left [ \delta \left ( k - 1 \right ) - M \left ( k - 1 \right ) \right ] mod 1 \\
\end{equation}
 
and

\begin{equation}\label{eq:19}
M(k) = \nicefrac{1}{N} + \nu (n)
\end{equation}

where $\nu (n)$ corresponds the loop filter output and $\delta(k)$ corresponds each sample calculated $\Delta$. The $ \nu (n)$ signal from the loop filter adjusts the amount by which the counter decrements and $\delta(k)$ is only passed to the interpolation filter as $\Delta$ if the strobe signal is valid. 

\subsection{Loop-Filter}

For both carrier and clock synchronization, a time discrete Proportional-Integral (PI) filter is used to calculate the control variable for the frequency and phase offset. The filter is a 1st order filter with the following transfer function in the z-domain \cite{Chung.1993}:

\begin{equation}\label{eq:20}
F\left ( z \right ) = \frac{C_2+C_1\left ( 1-z^{-1} \right )}{1-z^{-1}}
\end{equation}

The gain parameters $C_1$ and $C_2$ define the behavior of the closed loop and are calculated as follows:

\begin{equation}\label{eq:21}
C_0 = \frac{1}{K_0K_D} \cdot \frac{8\zeta \omega_nT_c}{4+4\zeta \omega_nT_c + \left ( \omega_nT_c \right )^2 }
\end{equation}

\begin{equation}\label{eq:22}
C_1 = \frac{1}{K_0K_D} \cdot \frac{4\left ( \omega_nT_c \right )^2}{4+4\zeta \omega_n + \left ( \omega_nT_c \right )^2 }
\end{equation}

$\zeta$ is the loop attenuation coefficient, $K_0$  the numerically-controlled oscillator (NCO) gain, $K_D$ the error detector gain, $T_c$ the chip interval and $\omega_n$  natural frequency. Values from \textbf{Table~\ref{table:2}} are used for the carrier synchronization loop respectively timing recovery loop. The NCO Gain $K_0$ correspond to the gradient of characteristic curve of the respective NCO. The Error Detector Gain $K_D$ can be taken from the respective S-curve of Error Detector \cite{Rice_2009}.

\begin{table}[htbp!]
\begin{center}
\begin{tabular}{r|cc}
      & \begin{tabular}[c]{@{}c@{}}Carrier\\ Loop\end{tabular} & \begin{tabular}[c]{@{}c@{}}Timing\\ Loop\end{tabular} \\ \hline
$K_0$       & 1                                                      & -1                                                    \\
$K_D$       & 1                                                      & 2,55                                                  \\
$\zeta$     & 0,7071                                                 & 0,7071                                                \\
$B_n$       & 200 Hz                                                 & 100 Hz                                                \\
$\omega_n$  & $1,89B_n$                                              & $1,89B_n$                                             \\
$T_c$       & $125\mu s$                                             & $125\mu s$                                                
\end{tabular}
\end{center}
\caption{Loops parameter}
\label{table:2}
\end{table}

\subsection{Frame Synchronisation}
\label{FrameSync}
After successful carrier synchronization and timing recovery, the next step is to locate a structure in the chip stream within which the valid data is located.

For the reliable detection of a frame start within the chip stream, a 78-chip synchronization word is used, which consists of a 64-chip long code word and a 14-chip long guard interval. The code words are lines of the Hadamard-Matrix $H_8$ spread with a spreading factor of 8, which are recursively derived from the formation rule:
\begin{equation}\label{eq:23}
H_{2n}=\begin{pmatrix}
H_n & H_n \\ 
H_n & -H_n
\end{pmatrix} 
\end{equation}

with

\begin{equation}\label{eq:22}
H_1 = 1 
\end{equation}

The selected codewords $(H_{8,3}, H_{8,5})$ have very good aperiodic autocorrelation properties. This ensures a sufficient distance between the maxima of frame start detection and the side lobes.

\begin{figure}[tbp!]
  \begin{center}
  \includegraphics[width=3.6in]{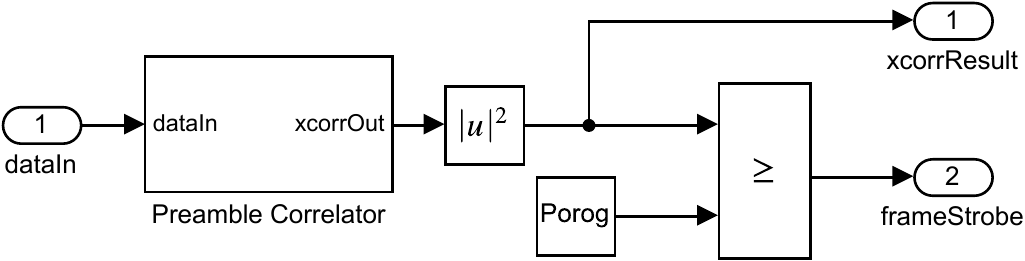}\\
  \caption{Frame Synchronisation Unit}
  \label{fig:9}
  \end{center}
\end{figure}

The detection of the synchronization pattern is performed by a correlator realized in FIR filter structure, which cross-correlates the local version of the synchronization pattern with the incoming baseband symbol stream. The amount of the cross-correlation is then squared. By applying a non-linear function, the distance between the main maxima and the possible sidelobes is further increased. This enables reliable detection even with a very low SNR. If the set threshold value is exceeded, a preamble is detected and valid data follows. The data can then be despreaded and demodulated. The block diagram of the frame synchronization unit is shown in \textbf{Figure~\ref{fig:9}}.

\subsection{Despreading and demodulation}
After synchronizing all three levels, the received data must now be despreaded and demodulated. After clock synchronization, chip stream to be treated is now available as chip/sample. The code phase position required for successful despreading of the data is also known due to the detection of the start of frames in the frame synchronization. Now the data is fed to a despreading unit in FIR filter structure. This delivers all L (length of code word) chips the despreaded symbol, which is converted into a data bit in the DBPSK demodulation unit. The resulting data bit stream is then stored in a queue register.


\section{Simulation results}
\label{sec:5}
\begin{figure}[tbp!]
  \begin{center}
  \includegraphics[width=0.6\linewidth]{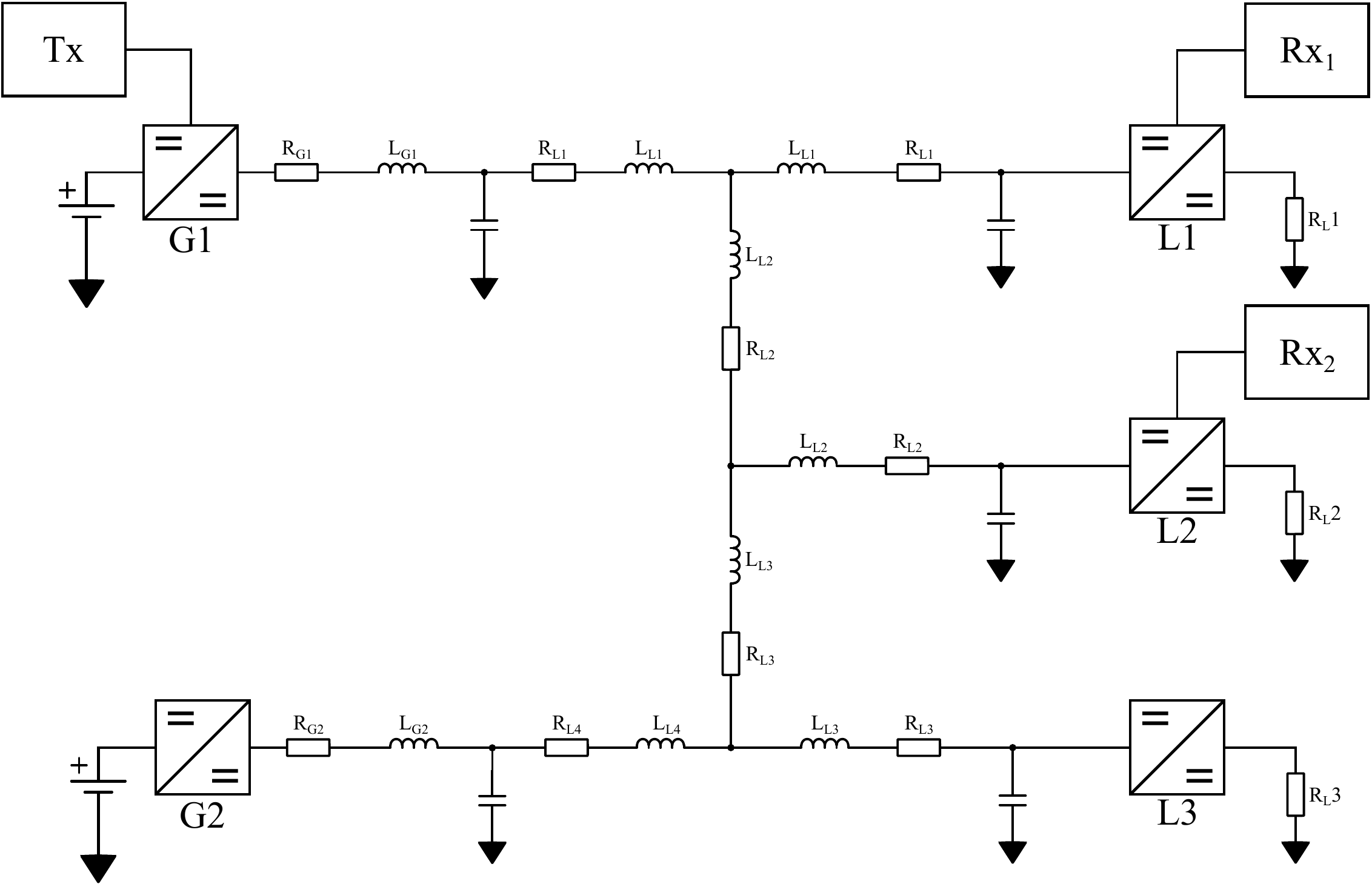}\\
  \caption{Simulated DC grid for testing the transmission system}
  \label{fig:s.5_p.1}
  \end{center}
\end{figure}

\textbf{Figure~\ref{fig:s.5_p.1}} shows the structure of the DC grid which was used for the simulative performance analysis of the implemented transmission system. It is a low voltage direct current circuit consisting of two generator converters (G1 and G2) and three consumer converters (L1-L3). Both the converters on the generator side and on the load side are synchronous DC-DC buck converters. This structure is based on the structure for testing the power/signal dual modulation (PSDM) Techniques for packet based energy dispatching from \cite{Din.2018}.  
This modified simulation example for the power/signal dual modulation technique, which in contrast to the original version is equipped with the transmission system developed in the present article, is intended to show that the information exchange between generator and load required for the exchange of the energy packets does not require a third-party synchronization unit. Furthermore, all parameters required for successful information and energy exchange can be derived from the received signal itself.

\begin{figure}[htbp!]
  \begin{center}
  \includegraphics[width=0.515\linewidth]{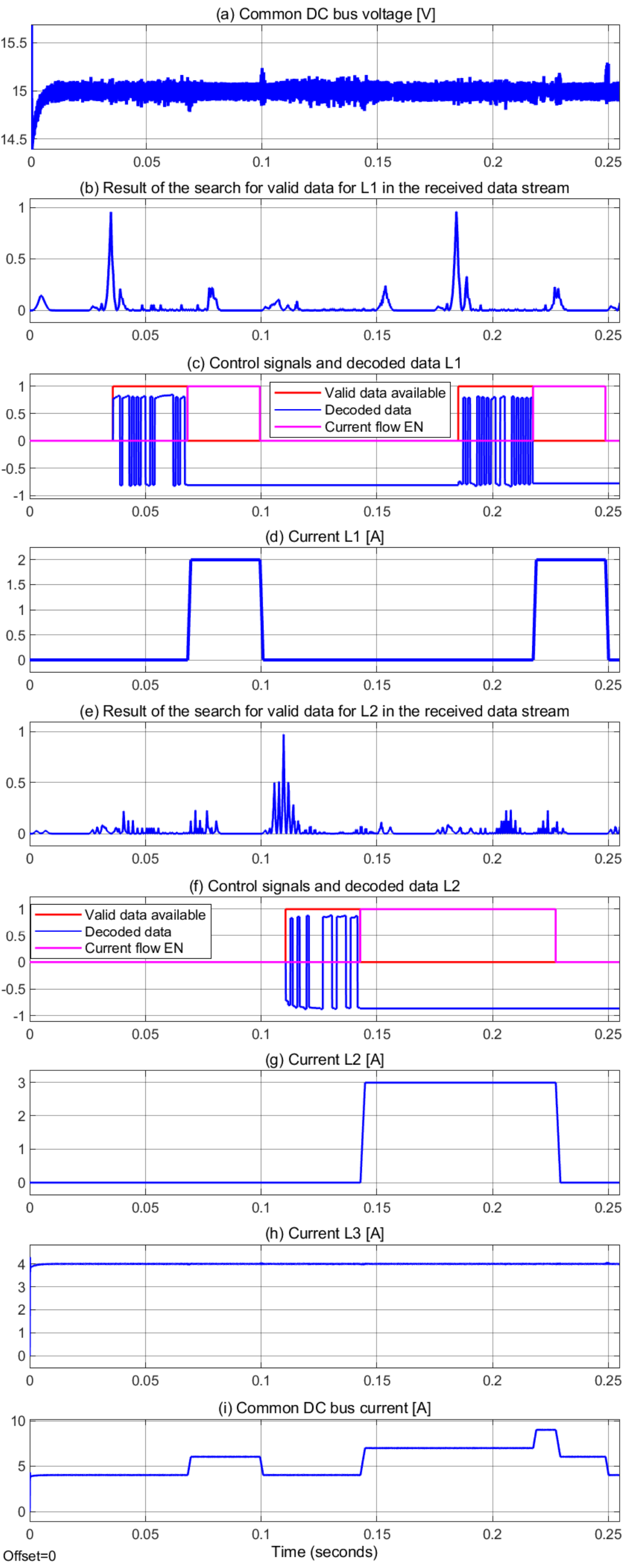}\\
  \caption{Simulation results}
  \label{fig:s.5_p.2}
  \end{center}
\end{figure}

\textbf{Figure~\ref{fig:s.5_p.2}} (a) shows the output signal of the generator (G1).  It supplies the DC bus with 15V DC voltage. In the context of power/signal dual modulation, the existing residual ripple of the DC voltage no longer represents only the unwanted noise; it is used constructively for information transmission. The messages sent by G1 for testing are coded alternately with the unique sequence S1 for load L1 and S2 for load L2. This DC voltage superimposed with the information signal is available as a broadcast signal to all devices connected to the bus.

At the beginning of this simulation example $t=0s$, the three loads are operated in listening mode. Simultaneously, L1 and L2 are configured as high-impedance $(I_{L1}=I_{L2}=0)$ and L3 is operated with a constant nominal current of $(I_{L3}=4 A)$, cf. Figure~\ref{fig:s.5_p.2} (d), (g) and (h).  As seen from the correlation peaks in Figure~\ref{fig:s.5_p.2} (b) L1 detects at time $t=35 \text{ ms}$  and $t=185 \text{ ms}$ data addressed to him within the received data stream.  After the synchronization, decoding and demodulation processes described in section~\ref{sec:4} have been performed, the receiver signals the presence of valid data by setting the signal "Valid data avaible" to high Figure~\ref{fig:s.5_p.2} (c) and passes the data to the control unit of load L1 for further processing. After elimination of the metadata by receiver, the data contains the control information regarding the amount of load current required and how long this should be present. This data is interpreted in the control unit and then the current flow enable signal is activated for the required time Figure~\ref{fig:s.5_p.2} (c) and the required current level is set for 35 ms. A similar process occurs at time t=110ms for load L2, see Figure~\ref{fig:s.5_p.2} (e), (f) and (g). The resulting total current is shown in subplot (i). In the following subsections the results of the realized synchronization level are presented. These enable the asynchronous transmission of energy packets via PSDM and derivation of metadata from the received data.

\begin{figure}[htbp!]
  \begin{center}
  \includegraphics[width=0.6\linewidth]{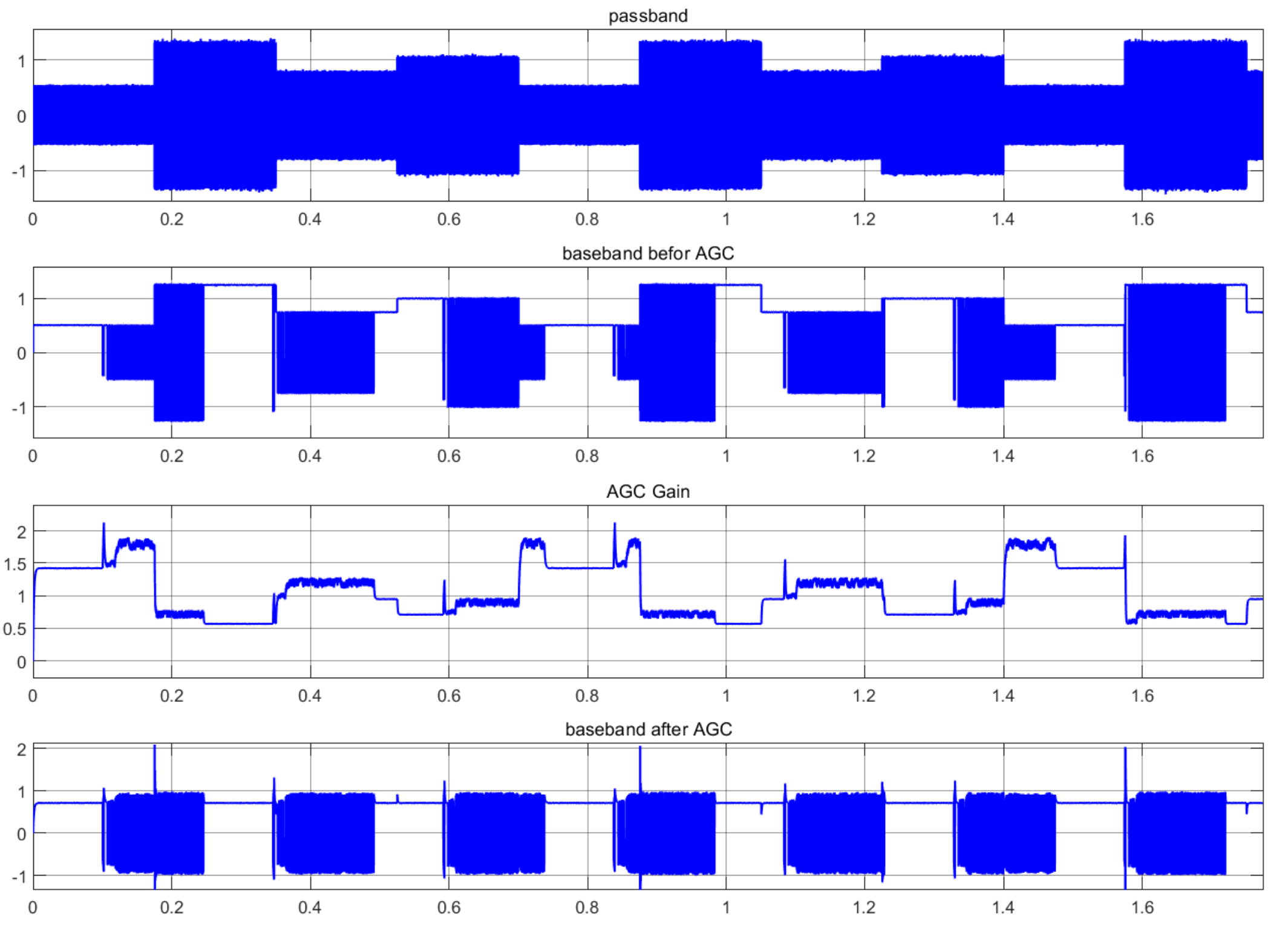}\\
  \caption{Results AGC test}
  \label{fig:s.5_p.3}
  \end{center}
\end{figure}

In order to test the AGC of the implemented receiver, the information part (i.e. high-frequency part) of the broadcast signal undergoes time-variable random amplification (G = 0.5 - 1.25). The change period is 175ms. \textbf{Figure~\ref{fig:s.5_p.3}} summarizes the simulation results of the Adaptive Gain Control tests. Subplot (a) shows the received signal in the passband, which varies randomly every 175ms. In subplot (b) the corresponding baseband signal before the AGC is shown, which changes in the same way. In subplot (c) the gain factor is shown which is applied to keep the output signal constant according to the reference. Finally, the subplot (d) shows the nearly constant baseband signal after the AGC.

\begin{figure}[htbp!]
  \begin{center}
  \includegraphics[width=0.6\linewidth]{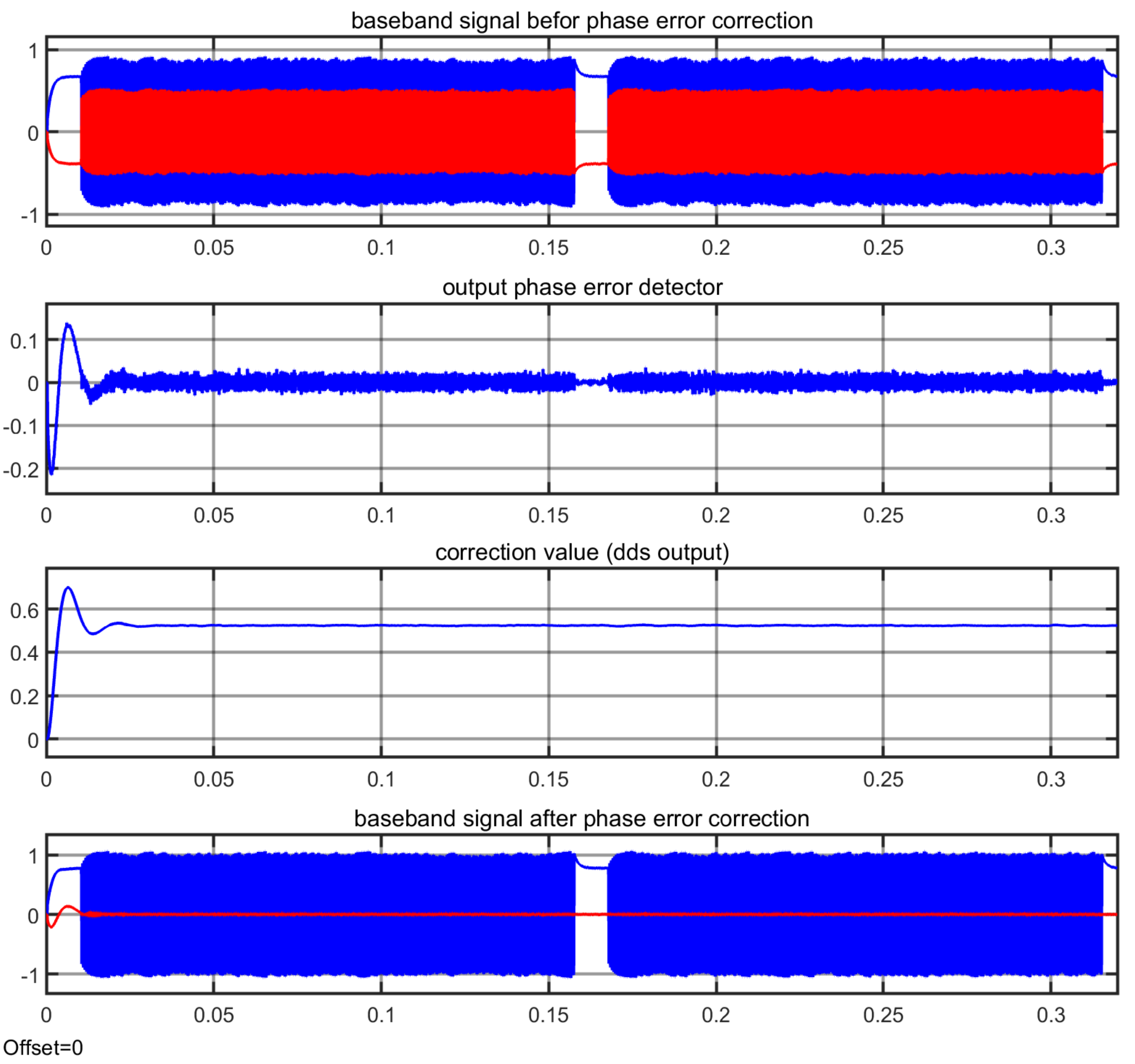}\\
  \caption{Results carrier synchronization for phase offset}
  \label{fig:s.5_p.4}
  \end{center}
\end{figure}

Next, the results of the phase and frequency offset correction are presented using the carrier synchronization unit. \textbf{Figure~\ref{fig:s.5_p.4}} shows the correction process of a phase offset between the carrier wave and the phase of the local oscillator of $\frac{\pi}{6}$. Subplot (a) shows the asynchronously downconverted baseband signal. Due to the phase shift, part of the signal power is transferred to the imaginary part (red). Subplot (b) shows the output of the phase error detector. Subplot (c) shows the resulting correction value $\widetilde{\phi}$ and subplot (d) the baseband signal after phase offset correction.

\begin{figure}[htbp!]
  \begin{center}
  \includegraphics[width=0.6\linewidth]{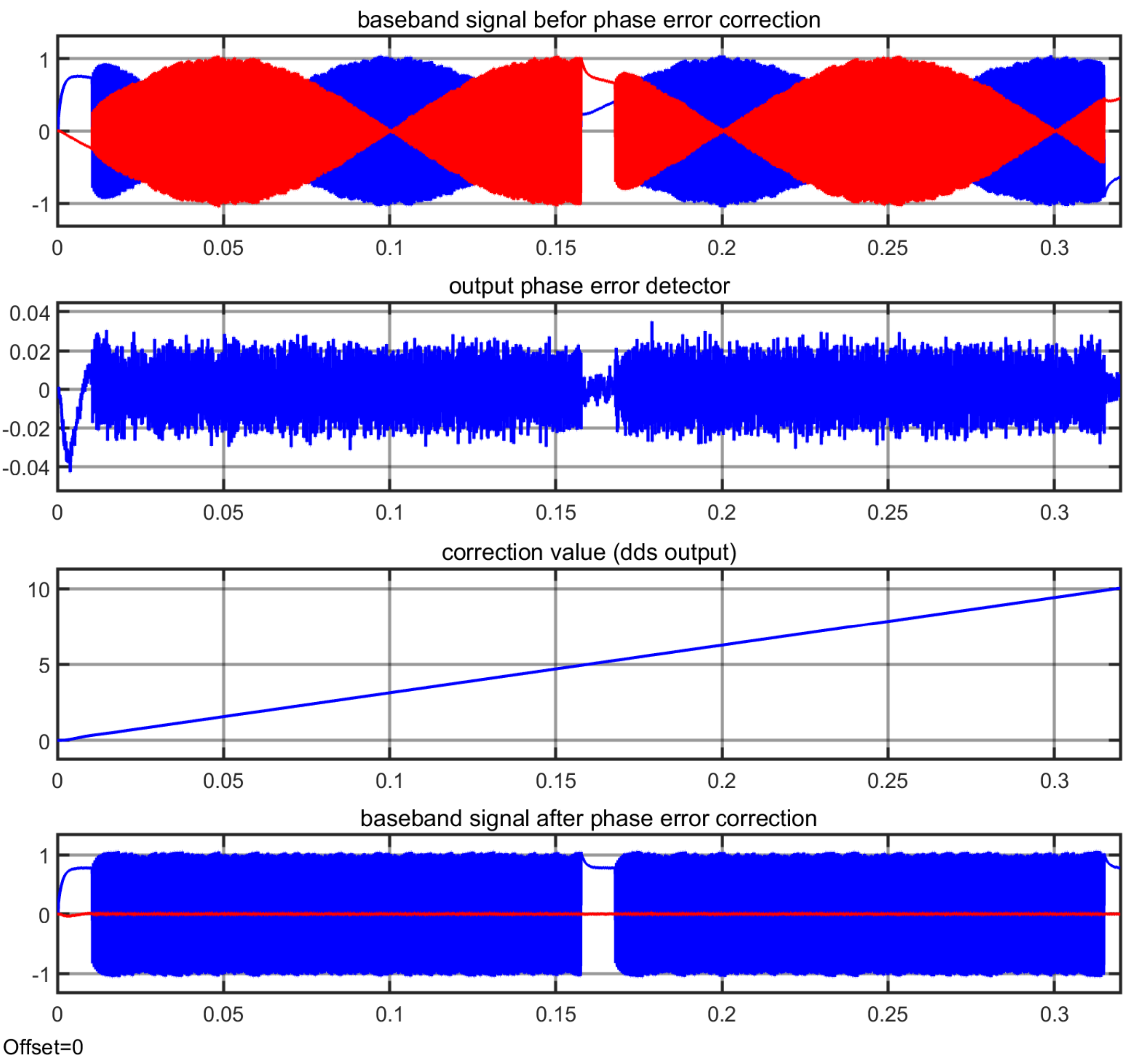}\\
  \caption{Results carrier synchronization for frequency offset}
  \label{fig:s.5_p.5}
  \end{center}
\end{figure}

The carrier synchronization results at a frequency offset are shown in \textbf{Figure~\ref{fig:s.5_p.5}}. The frequency offset that is used is $\Delta f = 5 Hz$.

\begin{figure}[htbp!]
  \begin{center}
  \includegraphics[width=0.6\linewidth]{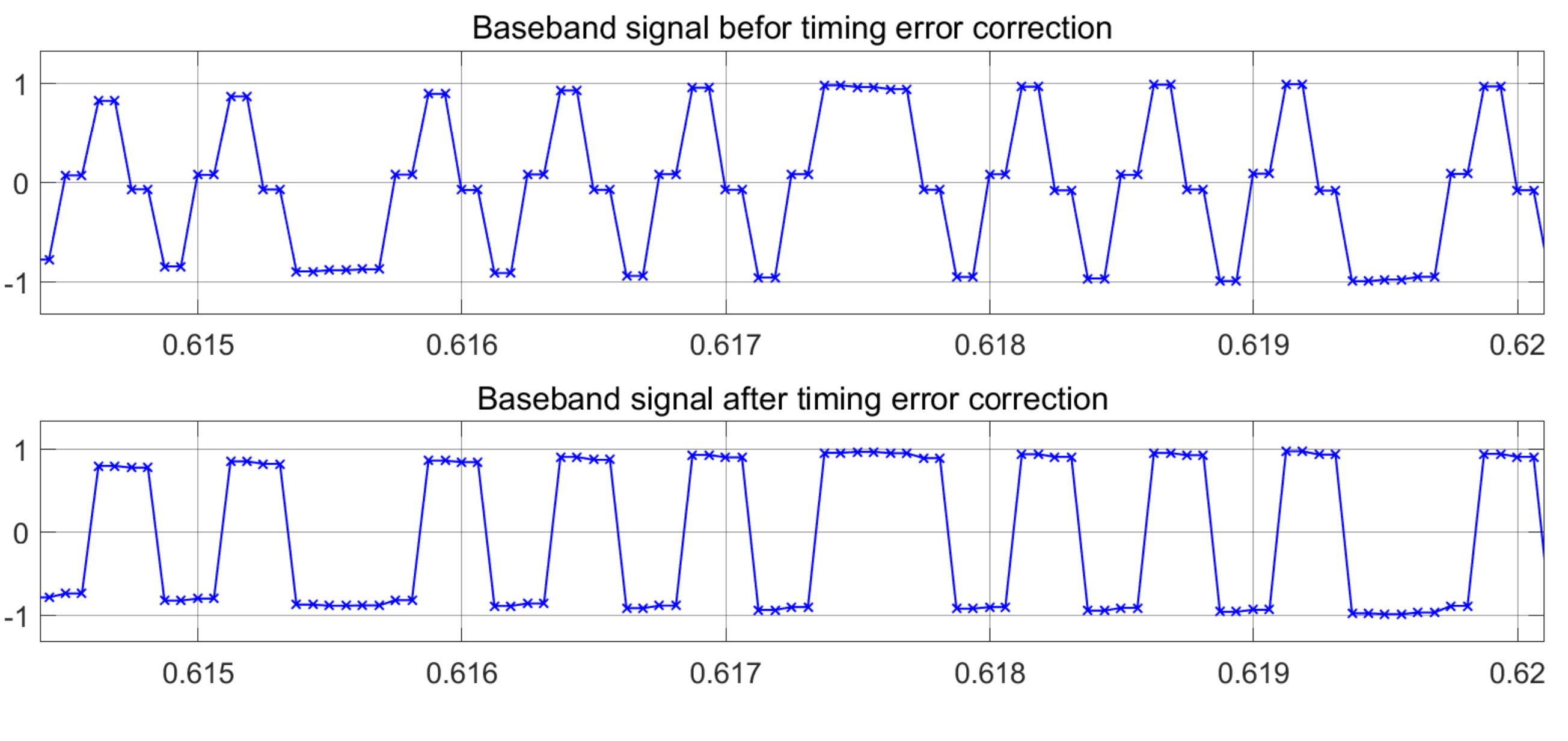}\\
  \caption{Baseband signal befor and after timming recovery}
  \label{fig:s.5_p.6}
  \end{center}
\end{figure}

\begin{figure}[htbp!]
    \centering
  \subfloat[\label{1c}]{%
        \includegraphics[width=0.3\linewidth]{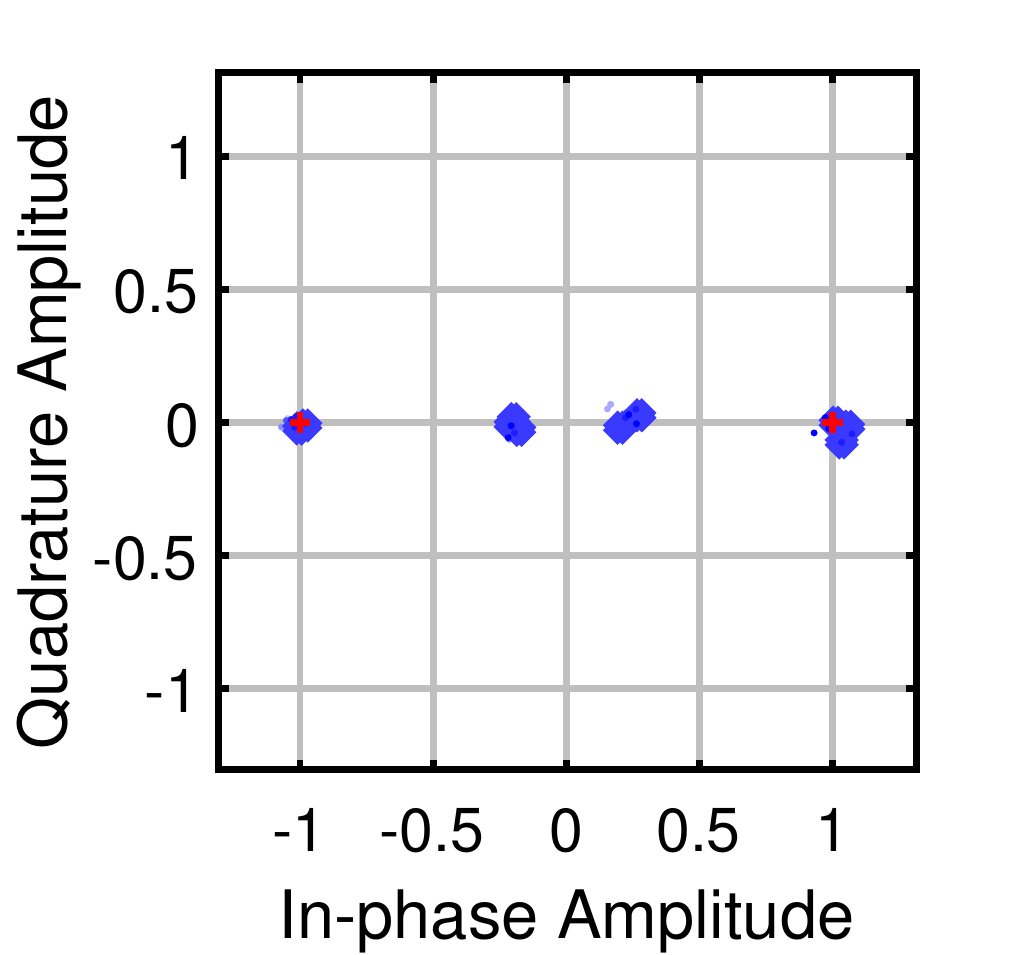}}
     \qquad
  \subfloat[\label{1d}]{%
        \includegraphics[width=0.3\linewidth]{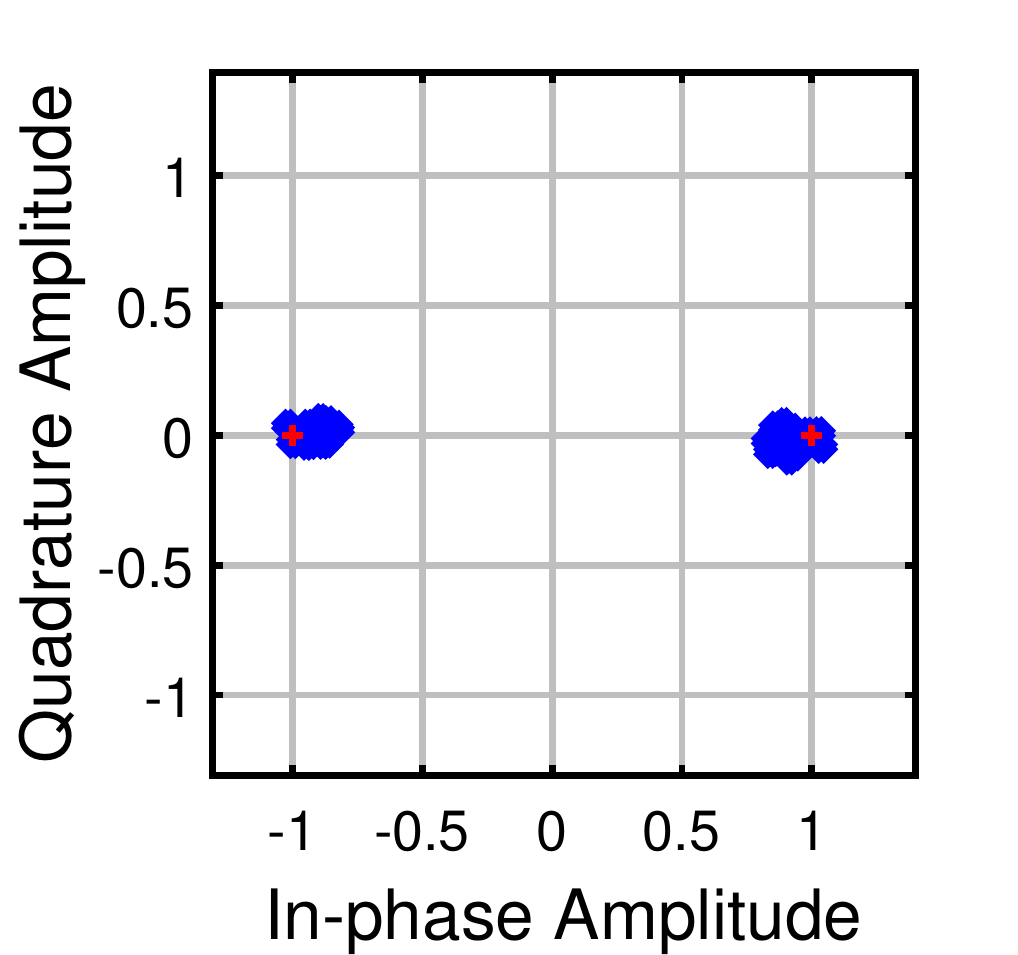}}
  \caption{Received signal constellation diagram. (a) befor and (b) after clock recovery}
  \label{fig:s.5_p.7} 
\end{figure}

\begin{figure}[htbp!]
  \begin{center}
  \includegraphics[width=0.6\linewidth]{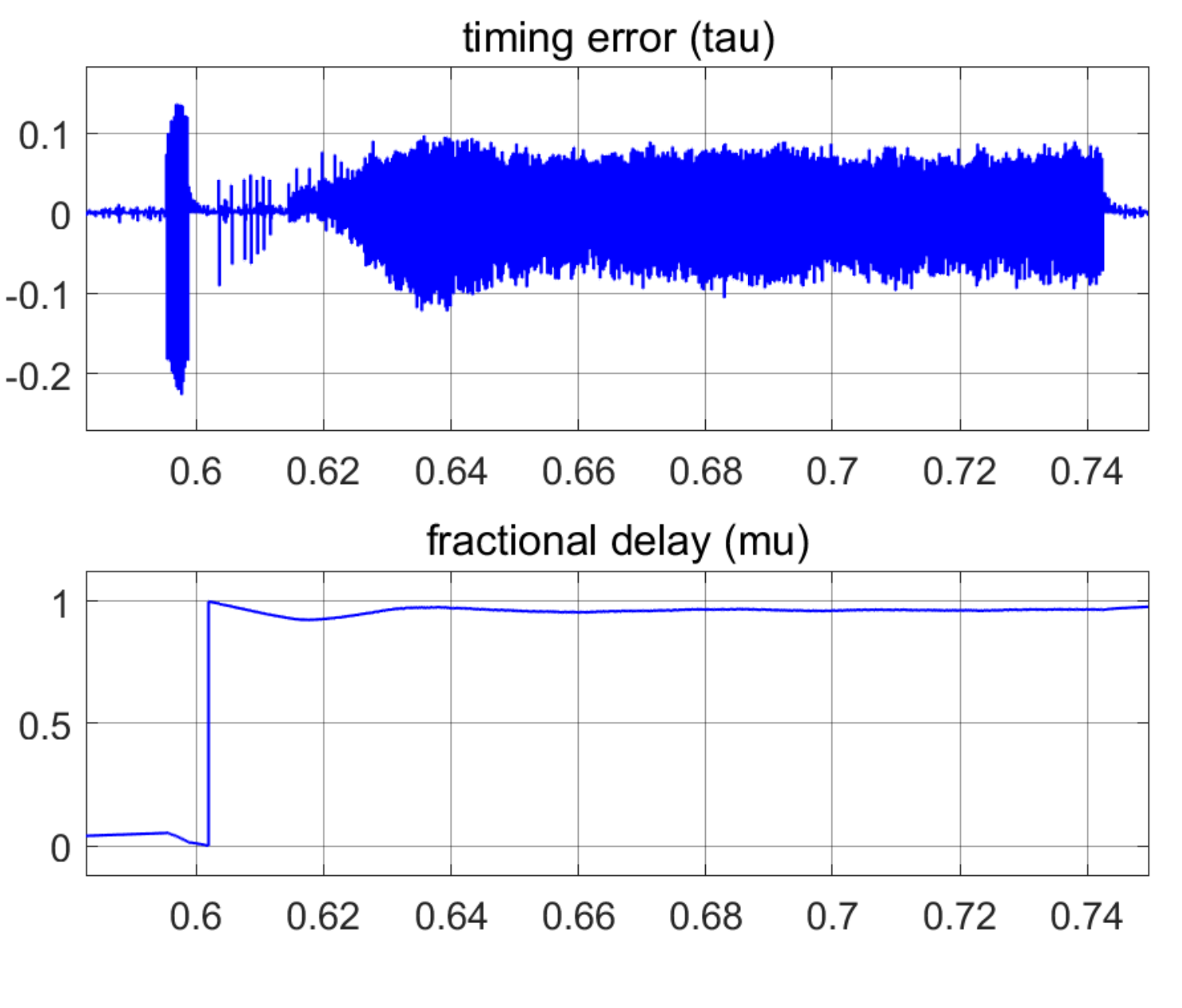}\\
  \caption{Estimated timing error and the resulting fractional delay}
  \label{fig:s.5_p.8}
  \end{center}
\end{figure}

The results of Chip Clock Recovery are summarized below. The sampling frequency offset of $\Delta f = 2 Hz$ is used to test the Timing Recovery Unit. Subplot (a) of \textbf{Figure~\ref{fig:s.5_p.6}} shows the bassisband signal of the received signal r(t) sampled without the local clock recovery. As one can see, plot (a) contains samples that should not be present in a DBPSK signal. These result from an asynchronous sampling. In the subplot (b) the same signal is shown but after the chip clock recovery. The constellation diagrams, Figure~\ref{fig:s.5_p.7}, further clarify the asynchronicity between the received data and the local clock as well as the result of the synchronisation by the clock recovery unit. \textbf{Figure~\ref{fig:s.5_p.8}} shows the corresponding estimated timing error and the resulting fractional delay for timing error correction.

\begin{figure}[htbp!]
  \begin{center}
  \includegraphics[width=0.6\linewidth]{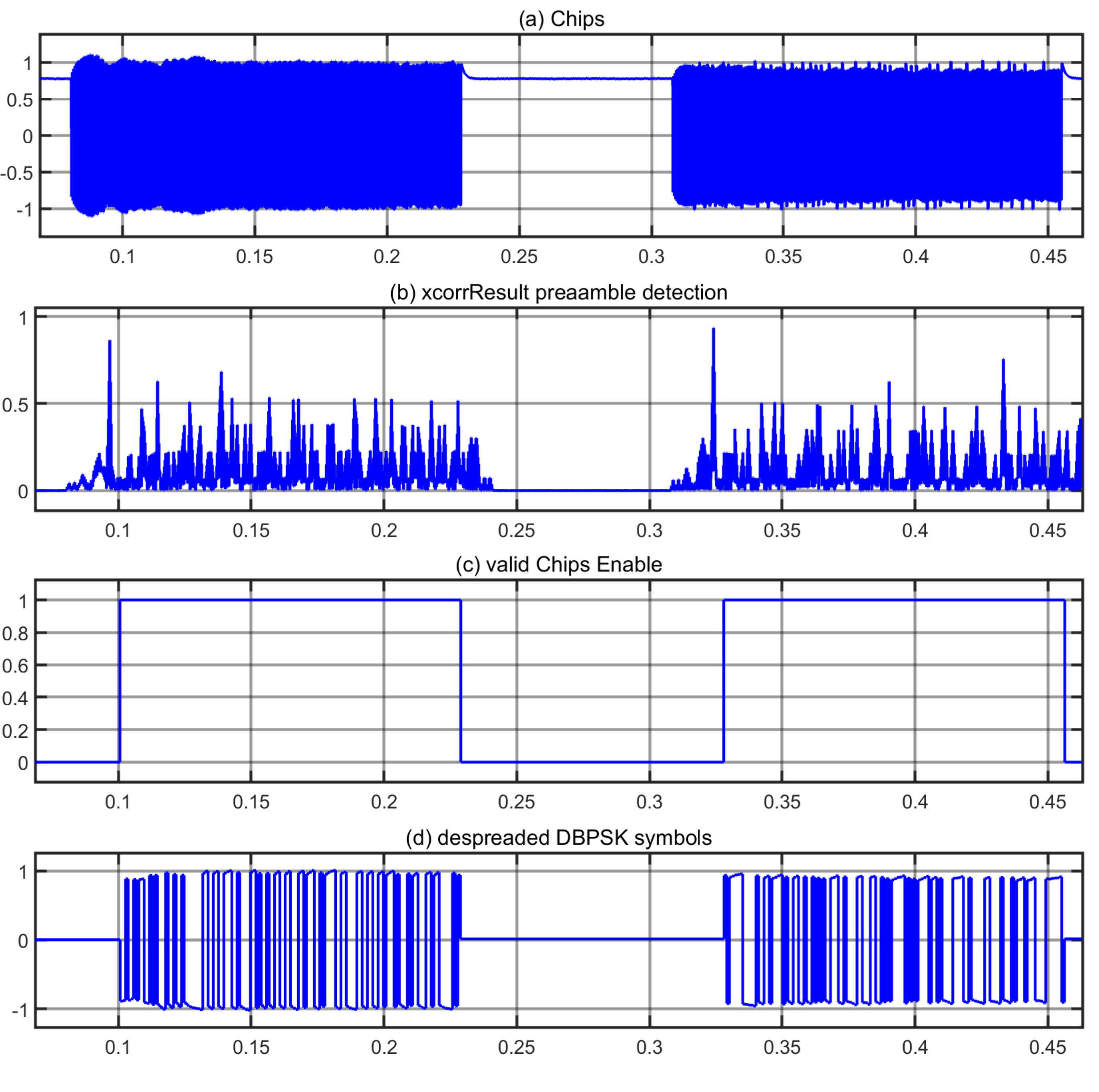}\\
  \caption{Result frame synchronisation load 1}
  \label{fig:s.5_p.9}
  \end{center}
\end{figure}

Finally, as described in Section~\ref{sec:4} , the frame synchronization, the despreading of the chips to the DBPSK symbols and finally the demodulation of the data is performed. The results of these operations for two received data frames is shown in \textbf{Figure~\ref{fig:s.5_p.9}}. Subplot (a) shows the received chipstream. The result of the cross correlation for the preamble detection is shown in the subplot (b). Subplot (c) shows the valid-frame-detected signal and subplot (d) shows the corresponding DBPSK symbols.

\begin{figure}[htbp!]
  \begin{center}
  \includegraphics[width=0.6\linewidth]{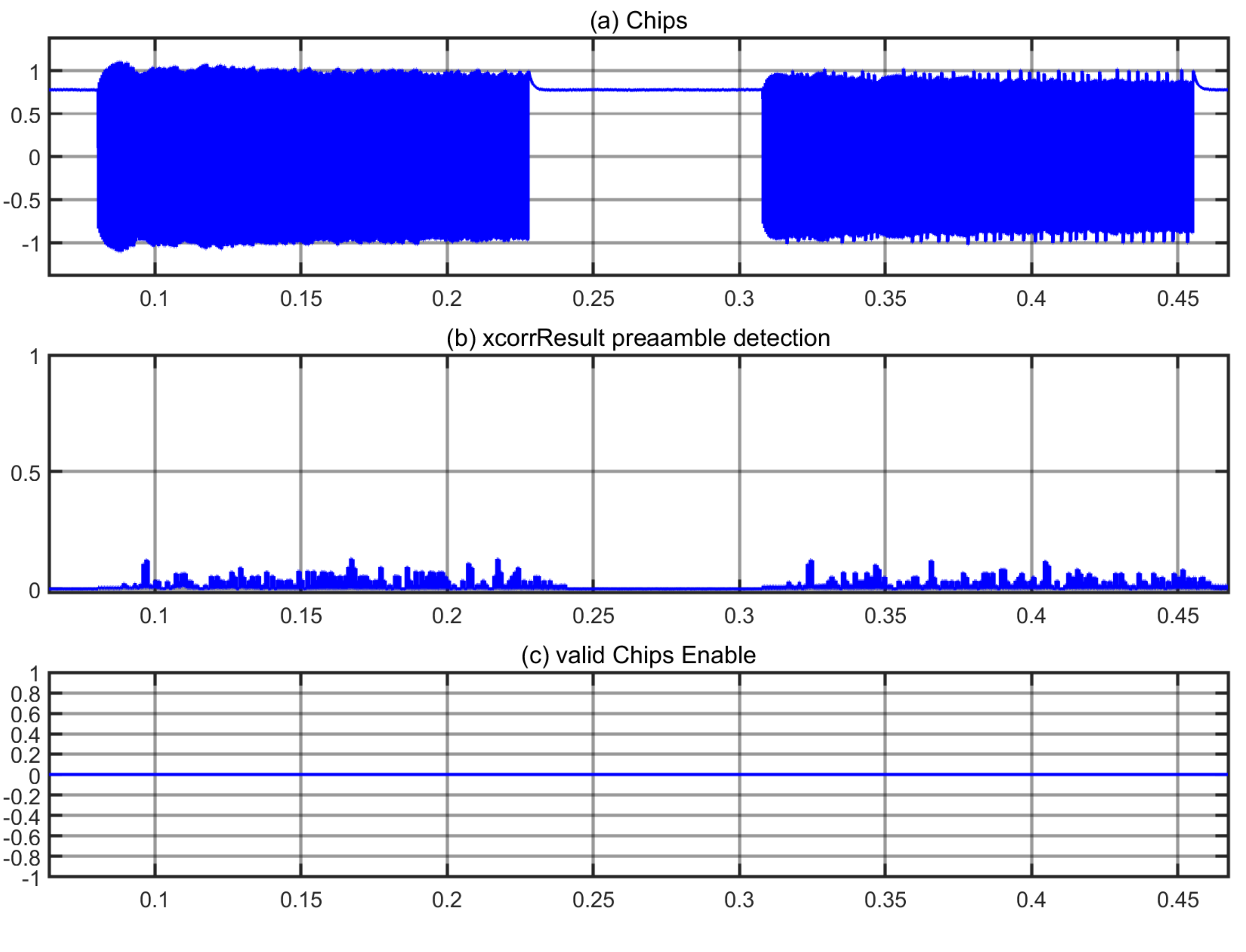}\\
  \caption{Result frame synchronisation load 2}
  \label{fig:s.5_p.10}
  \end{center}
\end{figure}

In contrast, \textbf{Figure~\ref{fig:s.5_p.10}} shows the same procedure but for the 2nd load. Since this load expects data coded with L2 unique sequence, no valid message is detected, which can be seen from the result of the cross correlation.

\begin{figure}[htbp!]
  \begin{center}
  \includegraphics[width=3in]{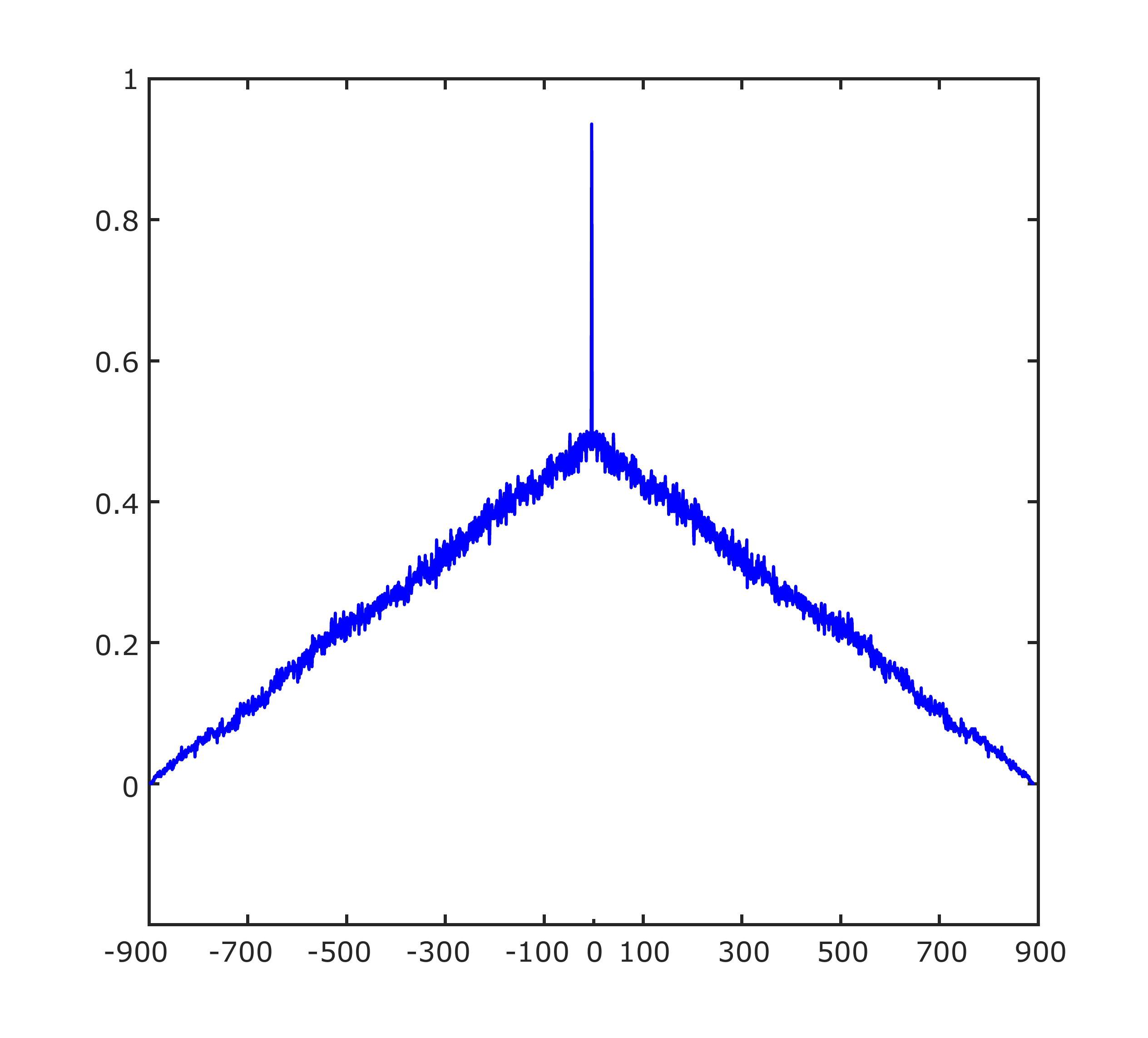}\\
  \caption{Result of the cross-correlation between sent and received data}
  \label{fig:s.5_p.11}
  \end{center}
\end{figure}

A series of ten information frames are sent to check the overall functionality of the implemented receiver. The received binary data are then cross-correlated with the transmitted data. The result of the cross-correlation is shown in \textbf{Figure~\ref{fig:s.5_p.11}}. A complete match of the transmitted and the received data can be recognized by the unique peak at $\tau=0$.

\section{Conclusion and Outlook}
\label{sec:6}

The present paper extends previous contributions \cite{8245741, Din.2018} regarding packet-based energy transmission by describing a practicable design of an energy packet receiver which recovers the required synchronicity information directly from the received signal itself. The main focus is on the reception and lossless preprocessing and interpretation of the metadata of the information part of the energy packet. For this purpose, implemented solutions for the respective synchronization levels (i) Carrier Recovery, (ii) Clock Recovery and (iii) Frame Recovery are discussed in detail. Drawing upon a DC grid example, simulation results show the performance and applicability of the proposed novel receiver for packet based energy dispatching. 

As future work, the proposed approach of packet-based energy distribution will be implemented and validated within the KIT Energy Lab 2.0 infrastructure \cite{Hagenmeyer.2016}.

\medskip

%
\bibliographystyle{MSP}
\bibliography{Bibliography}

\end{document}